\documentclass[nojss,shortnames]{jss}

\usepackage{amsfonts}
\usepackage{tikz}
\usetikzlibrary{shapes.geometric, arrows, positioning}
\tikzstyle{io} = [rectangle, text width=5em,text centered, draw=black]
\tikzstyle{ana} = [rectangle, text width=5em,text centered, draw=black]
\tikzstyle{sum} = [rectangle, text width=7em,text centered, draw=black]
\tikzstyle{vis} = [rectangle, text width=9.4em,text centered, draw=black]
\tikzstyle{title} = [rectangle, text width=5em, text centered]
\tikzstyle{arrow} = [thick,->,>=stealth]

\usepackage{orcidlink,thumbpdf,lmodern,amsmath,amssymb}
\usepackage{hyperref}
\usepackage{graphicx}
\usepackage{subcaption}
\usepackage{algorithm}
\usepackage{algorithmic}
\usepackage{framed}
\usepackage{float}

\author{Zhuoran Yu \\ Emory University
   \And Armin Schwartzman \\ University of California San Diego
   \AND Junting Ren \\ University of California San Diego
   \And Julia Wrobel \\ Emory University}
\Plainauthor{Zhuoran Yu, Armin Schwartzman, Junting Ren, Julia Wrobel}

\title{\pkg{SCoRES}: An \proglang{R} Package for Simultaneous Confidence Region Estimates}
\Plaintitle{SCoRES: An R Package for Simultaneous Confidence Region Estimates}
\Shorttitle{SCoRES for Simultaneous Confidence Region Estimates}

\Abstract{
The identification of domain sets whose outcomes belong to predefined subsets can address fundamental risk assessment challenges in climatology and medicine. Existing approaches for inverse domain estimates
require restrictive assumptions, including domain density and continuity of function near thresholds, and large-sample guarantees, which limit the applicability. Besides, the estimation and coverage depend on setting a fixed threshold level, which is difficult to determine. Recently, \cite{Ren:2024} proved that
 confidence sets of multiple levels can be simultaneously constructed with the desired confidence
 non-asymptotically through inverting simultaneous confidence bands. Here, we present the \textbf{SCoRES} R package, which implements Ren's approach for both the estimation of the inverse region and the corresponding simultaneous outer and inner confidence regions, along with visualization tools. Besides, the package also provides functions that help construct SCBs for regression data, functional data and geographical data.
To illustrate its broad applicability, we present three rigorous examples that demonstrate the \textbf{SCoRES} workflow.
}

\Plainkeywords{JSS, style guide, comma-separated, not capitalized, R}

\Address{
  Julia Wrobel\\
  Emory University\\
  Department of Biostatistics and Bioinformatics\\
  1518 Clifton Rd, Atlanta\\
  Georgia 30322, United States of America\\
  E-mail: \email{julia.wrobel@emory.edu}
}

\begin{document}

\section{Introduction}


In many statistical applications, researchers are not only interested in the value of a function itself, but in where that function exceeds (or falls below) a scientifically meaningful threshold. These regions are called inverse sets (or excursion sets), because they consist of all points in the domain where the underlying function crosses the threshold. For example, in climate science one may want to identify the parts of North America where the projected increase in temperature exceeds 2°C, or in regression settings, to identify which patient profiles correspond to a risk greater than 40\%. In such problems, it is crucial not only to estimate which region exceeds the threshold, but also to provide a measure of uncertainty around that region — just as confidence intervals quantify uncertainty for numerical parameters. This leads to confidence sets (CSs) for inverse sets: upper and lower bounds that enclose, with statistical guarantee, the true excursion region.

A general theoretical framework for constructing confidence bounds for inverse sets was introduced by \cite{Mammen:2013}, and since then inverse set estimation has been applied in several scientific domains, including astronomy \citep{Jang:2006}, medical imaging \citep{Bowring:2019, Bowring:2021, WillettNowak:2005}, dose-effect finding \citep{jankowski:2014}, geoscience \citep{french:2017}, and climate change \citep{Sommerfeld:2018}. However, most existing approaches focus either on dense functions \citep{saavedra:2016, qiao:2019} or on random functions arising from stochastic processes \citep{french:2013, french:2014, bolin:2015}. These methods typically provide valid confidence sets only for a single fixed threshold $c$, and do not offer simultaneous control across multiple thresholds, which limits their applicability for exploratory or data-driven analysis.

While simultaneous confidence bands (SCBs) quantify uncertainty for the \textit{values} of a function, they do not directly address uncertainty about the \textit{regions} where the function exceeds a threshold. For that purpose, one needs simultaneous confidence regions (SCRs) for the corresponding inverse set. The \textbf{SCoRES} method adopts the recent approach of \cite{Ren:2024}, which constructs SCRs by inverting SCBs. Because SCBs are widely available across statistical models, this inversion framework is broadly applicable — extending beyond dense functional data to regression models and other commonly encountered settings. Moreover, the resulting SCRs achieve simultaneous coverage with probability at least $1-\alpha$ across all thresholds under consideration. This enables users to explore multiple thresholds sequentially, without pre-specification, while retaining strict Type~I error control. In the following examples, we illustrate why this property and SCBs more broadly are particularly useful in practice.

As a first illustration, consider a study of pupil response to light following cannabis use. Pupillary light reflex has recently been investigated as a potential digital biomarker of recent cannabis use, motivated by the need for objective roadside or workplace impairment assessments when blood THC levels are unreliable \citep{godbole2024study}. In this study, participants with daily, occasional, or no cannabis use histories completed a standardized light response test, and functional data analysis revealed systematic differences in the degree of pupil constriction between recent users and non-users \citep{godbole2024study}. Suppose we are interested specifically in the relative change in pupil size over time between users and non-users. A functional-on-scalar regression (FoSR) model can be used to estimate this difference:

 $$Y_i(t) = \beta_0(t) + \beta_1(t) \mathbf{1}\{\text{use}_i=1 \} + b_i(t) + \epsilon_i(t),$$ 
 
\noindent
where $Y_i(t)$ is the relative change in pupil size at time $t$ for the $i$th subject, and $\beta_1(t)$ represents the time-varying group difference. 

\begin{figure}[H]
\centering

    \includegraphics[width=\linewidth]{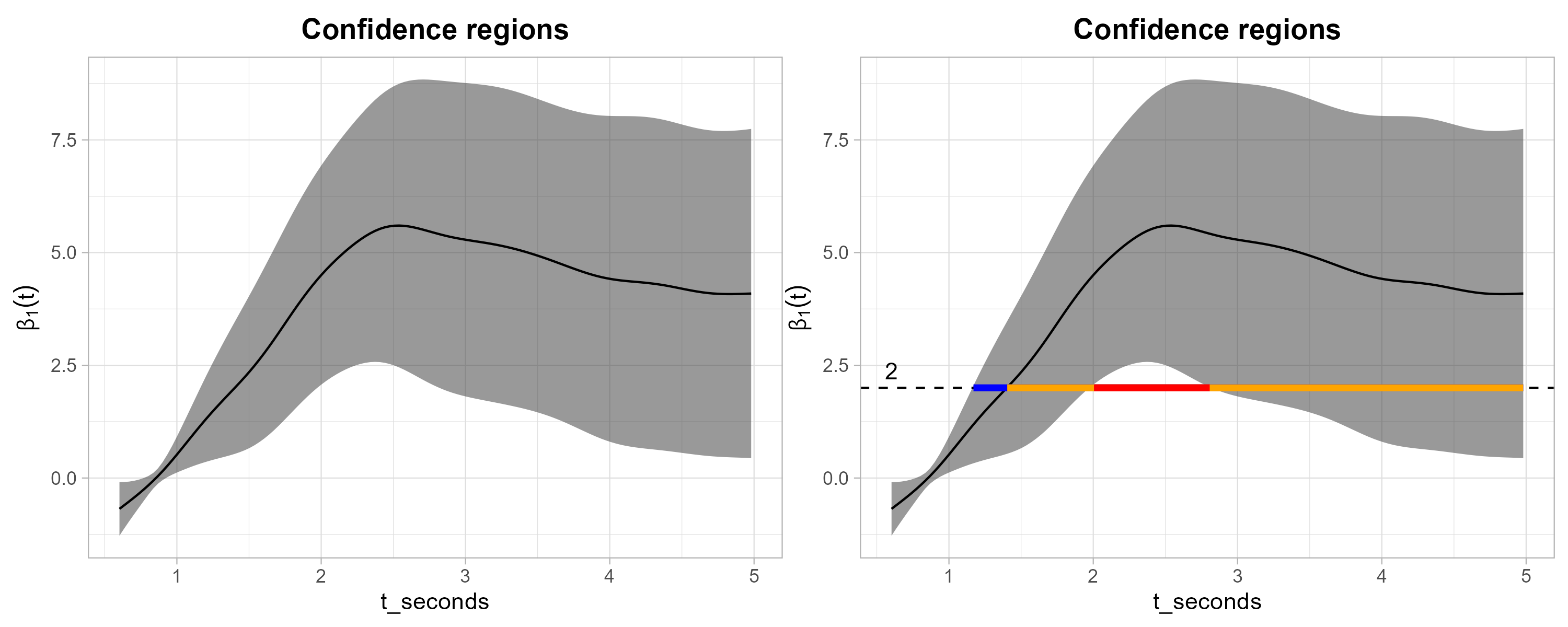}
\caption{
\textbf{Simultaneous inference for functional outcomes.}
Left: Simultaneous confidence bands (SCBs) show the pointwise uncertainty for a functional estimate.
Right: Simultaneous confidence regions (SCRs) visualize the joint uncertainty region across the domain.
The yellow+red region shows the point estimate of the excursion set, the red region is the 95\% inner confidence set, and the blue+yellow+red region is the 95\% outer confidence set.
}
\label{fig:scb_scr}
\end{figure}

The left panel of Figure \ref{fig:scb_scr} shows a simultaneous confidence band (SCB) for $\beta_1(t)$, which quantifies uncertainty in the estimated trajectory. However, in many applied settings the scientific question is not whether the curves differ, but when and for how long the effect exceeds a meaningful threshold, such as $c = 2$.

The right panel of Figure~\ref{fig:scb_scr} displays the corresponding simultaneous confidence region (SCR) for the inverse set ${t : \beta_1(t) \ge 2}$, constructed using \textbf{SCoRES}. The yellow + red segment gives the point estimate of the time interval where the effect exceeds 2, the red region represents the 95\% inner confidence set (where exceedance is supported with high confidence), and the blue + yellow + red region forms the 95\% outer confidence set. From this, we conclude that the difference in pupil response reliably exceeds the threshold beginning at approximately 1.5 seconds after stimulus onset.

A second example comes from climatology, where researchers are interested in how much average summer temperature has increased across North America between the late 20th century and the mid-21st century. A common approach is to fit a generalized least squares (GLS) regression model of the form

$$\text{Temperature} = \beta_0 + \beta_1 \cdot \text{Group} + \beta_2 \cdot \text{Time}_{\text{current}} + \beta_3 \cdot \text{Time}_{\text{future}} + \epsilon,$$ 

\noindent
where \text{Group = 1} denotes future years (2041–2069) and \text{Group = 0 denotes historical }\\ \text{observations (1971–1999).} The parameter $\beta_1$ captures the projected change in mean summer temperature between these two periods. Although a multiplier-t bootstrap can be used to construct a simultaneous confidence band (SCB) for $\beta_1$ across the spatial domain, this still leaves two practically important questions unanswered:
(1) which geographic regions are expected to experience warming above a chosen policy-relevant threshold (e.g., 2°C), and (2) with what spatial uncertainty can we assert that those regions will exceed the threshold?

\begin{figure}[H]
    \centering
        \includegraphics[height=10cm, width=\linewidth, keepaspectratio]{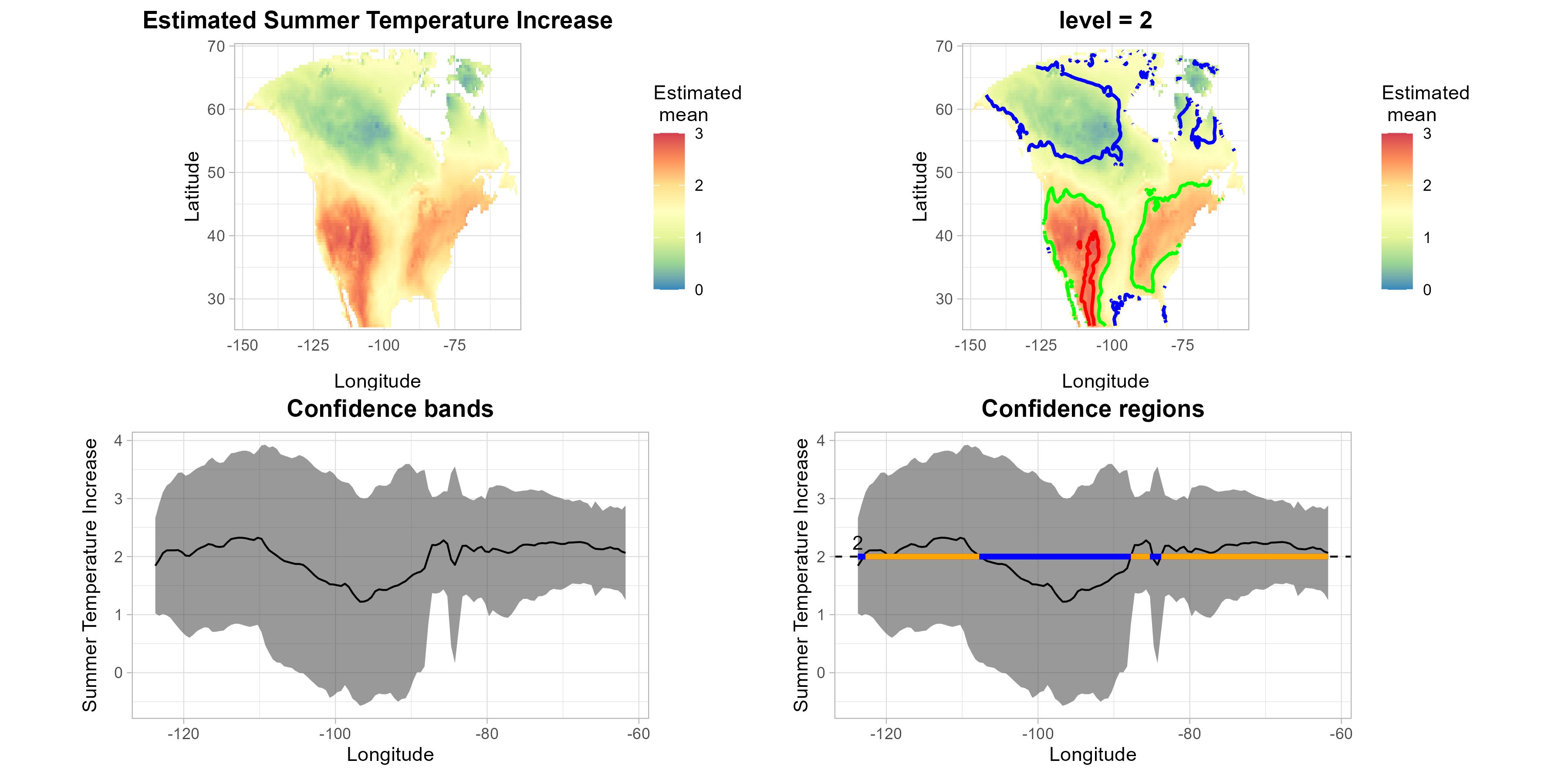} 
        \caption{
\textbf{Simultaneous inference for increase in summer temperature across North America.}
Top Left: Estimated summer temperature increase across North America.
Top Right: Simultaneous confidence regions (SCRs) visualize the joint uncertainty region across the 2D domain.
The green region shows the point estimate of the excursion set, the red region is the 90\% inner confidence set, and the blue region is the 90\% outer confidence set.
Bottom Left: Simultaneous confidence bands (SCBs) show the pointwise uncertainty for summer temperature increase estimate across the longitude at 40 Latitude.
Bottom Right: Simultaneous confidence regions (SCRs) visualize the joint uncertainty region across the longitude at 40 latitude.
The yellow region shows the point estimate of the excursion set, and the blue+yellow region is the 90\% outer confidence set. The 90\% inner confidence set here is an empty set.}
        \label{fig:geo-SCR}
\end{figure}

Figure~\ref{fig:geo-SCR} shows the estimated increase in summer temperature across North America alongside the SCB for a fixed latitude of 40°. As in the previous example, the SCB quantifies uncertainty in the \textit{magnitude} of $\beta_1$, but does not directly reveal \textit{which} spatial locations are likely to exceed the threshold. The simultaneous confidence region (SCR) provides this missing information: the green contour represents the point estimate of the region where warming exceeds 2°C, the red contour shows the 90\% inner confidence region (where exceedance is supported with high confidence), and the blue contour shows the 90\% outer confidence region (the set of locations where exceedance cannot be ruled out). Together, these contours give a spatially explicit and statistically valid characterization of both the extent of warming and the uncertainty associated with it.

In addition, we consider a third example based on the \texttt{bp\_preg} dataset from package \textbf{bp} by \citep{bp:2022}, which records repeated blood pressure measurements collected every 30 minutes during the Pregnancy Day Assessment Clinic (PDAC) observation period for up to a maximum of 240 minutes among 209 pregnant women. This recent investigation in obstetrics and gynecology examined predictors of pregnancy-induced hypertension (PIH) and pre-eclampsia (PE) using cardiac and physiological measurements, aiming to evaluate whether blood pressure readings obtained within the first 1-hour (60-minute) observation period were adequate compared with the conventional 4-hour (240-minute) assessment window.

In this analysis, we are interested in assessing whether renal impairment, as reflected by elevated serum creatinine, is associated with higher systolic blood pressure (SBP). To summarize repeated measurements for each participant, we compute the average SBP across all time points and fit a simple linear regression model:

\[
\text{SBP}_{i} = \beta_{0} + \beta_{1}\,\text{Creatinine}_{i} + \varepsilon_{i},
\]

where $\text{SBP}_{i}$ represents the mean systolic blood pressure for participant $i$, and $\text{Creatinine}_{i}$ denotes the subject's serum creatinine level measured in $\mu\text{mol}/\text{L}$. The coefficient $\beta_{1}$ quantifies the expected change in SBP per unit increase in creatinine. An interesting research question would be to determine the creatinine threshold above which systolic blood pressure (SBP) is likely to exceed a clinically meaningful level, such as 130 mmHg — a commonly used cutoff indicating elevated or borderline hypertension in pregnancy.

\begin{figure}[H]
    \centering
        \includegraphics[height=10cm, width=\linewidth, keepaspectratio]{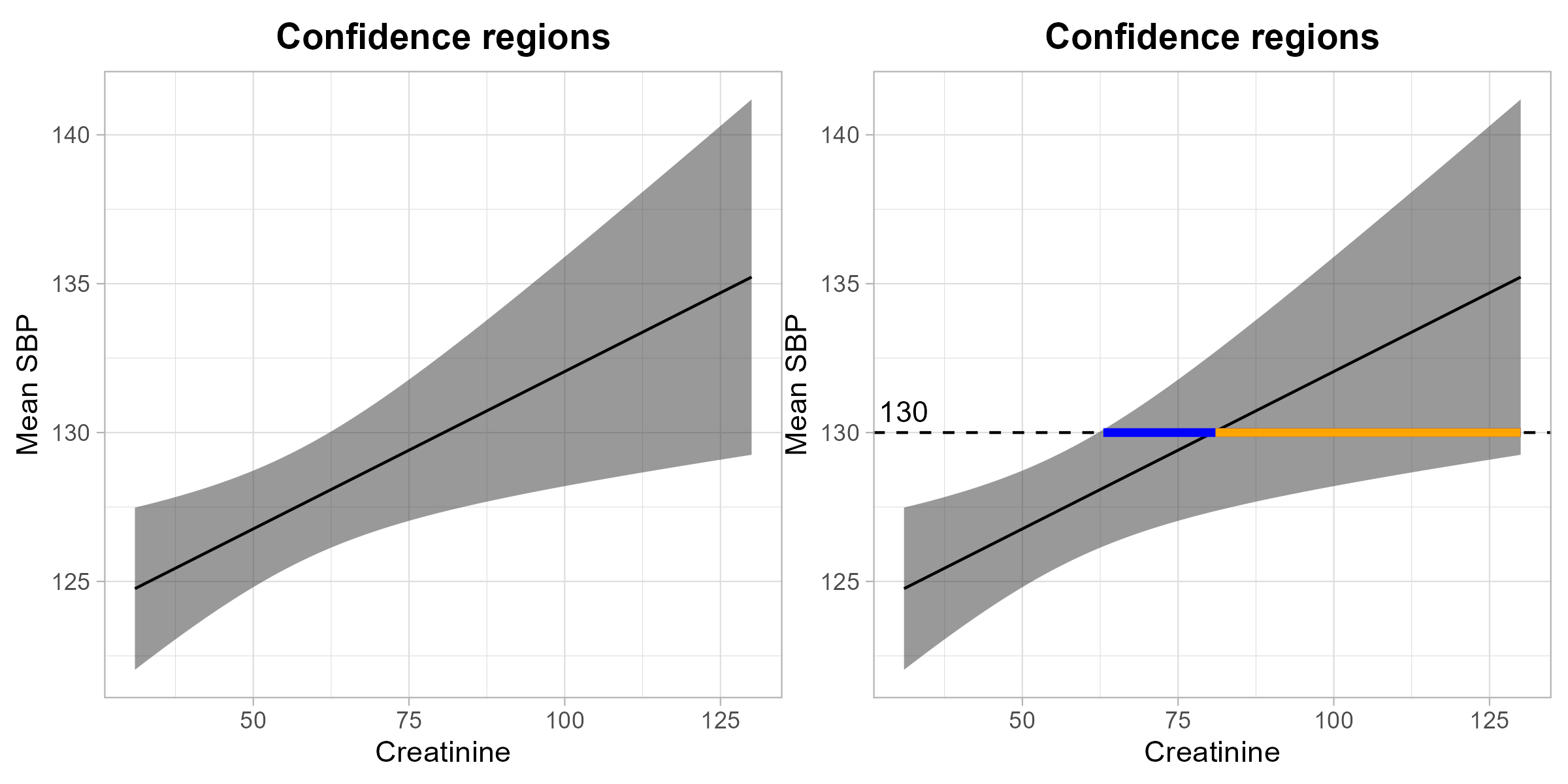} 
        \caption{
\textbf{Simultaneous inference for predicted SBP over creatinine's levels.}
Left: Simultaneous confidence bands (SCBs) show the pointwise uncertainty for predicted SBP over creatinine's levels.
Right: Simultaneous confidence regions (SCRs) visualize the joint uncertainty region across the domain.
The yellow region shows the point estimate of the excursion set, and the blue+yellow region is the 95\% outer confidence set. The 95\% inner confidence set here is an empty set.}
        \label{fig:creatinine_sbp}
\end{figure}

The left panel of Figure~\ref{fig:creatinine_sbp} shows a simultaneous confidence band (SCB) for predicted $\text{SBP}_{i}$ over $\text{Creatinine}_{i}$ ranging from 31-130, which quantifies uncertainty in the prediction. However, in many applied settings the clinical question is not about how to predict SBP, but to understand to what extent an elevated creatinine level would make a patient more likely to have a systolic blood pressure (SBP) exceeding 130 mmHg.

The right panel of Figure \ref{fig:creatinine_sbp} displays the corresponding simultaneous confidence region (SCR) for the inverse set ${Creatinine : \text{SBP}_{i} \ge 130}$, constructed using \textbf{SCoRES}. The yellow segment gives the point estimate of the time interval where the predicted SBP exceeds 130, and the blue + yellow region forms the 95\% outer confidence set. In this case, the 95\% inner confidence set is an empty set. From this, we conclude that if a patient has a measurement of creatinine around 80, then she would be likely to have a SBP above 130.

Our software package \textbf{SCoRES} provides a general and extensible framework for estimating SCRs by inverting SCBs, with rigorous simultaneous Type I error control. It supports upper-excursion, lower-excursion, and interval sets, and includes built-in visualization tools for discrete, one-dimensional, and two-dimensional domains. These visualizations help researchers interpret regional uncertainty in a principled way, even in complex spatial or functional settings.

Because the validity of SCRs fundamentally depends on the quality of the underlying SCBs, the performance of \textbf{SCoRES} is closely tied to accurate SCB estimation. However, no existing software package provides SCB construction across the broad range of model classes commonly encountered in practice, including linear and logistic regression, spatial GLS, and functional-on-scalar regression (FoSR). This is particularly true for functional data settings, which have seen limited software support. To address this gap, \textbf{SCoRES} implements several approaches for SCB estimation, including the nonparametric bootstrap, the multiplier-t bootstrap, and a parametric simulation approach for functional data, the Correlation and Multiplicity Adjusted (CMA) method \citep{crainiceanu2024functional}. We further evaluate these methods through simulation studies, comparing their empirical coverage properties across multiple data-generating regimes.

In this paper, we present both the theoretical foundation underlying the \textbf{SCoRES} framework and the software implementation for constructing SCBs and SCRs. We also provide an overview of the package structure, main functions, and software dependencies, and illustrate the full workflow through three detailed examples spanning functional, spatial, and regression settings. The remainder of this paper is organized as follows: Section~\ref{sec:methods} introduces the methodology for SCB and SCR construction; Section~\ref{sec:software} describes the structure and implementation of the \textbf{SCoRES} package; Section~\ref{sec:examples} presents the applied examples; and Section~\ref{sec:discussion} concludes with remarks on practical considerations, limitations, and future extensions.
 
\section{Statistical Methods}
\label{sec:methods}
  \subsection{SCoRES Definition}

  SCoRES (Simultaneous Confidence Region Excursion Sets) is defined as the estimate of the simultaneous confidence region in the context of Coverage Probability Excursion (CoPE) sets from \cite{Sommerfeld:2018}. It aims to assess the spatial
  uncertainty of domain sets whose outcomes belong to predefined subsets of the real line. A full introduction to the SCoRES concept can be found in previous work from \cite{Ren:2024}.

  Mathematically, the target region that corresponds to the inverse image of $U \subset  \mathbb{R}$ under an unknown function $\mu: \mathcal{S} \to \mathbb{R}$, can be defined as 
  $\mu^{-1}(U) = \{s \in S: \mu(s) \in U\}$
  , with $U$ a pre-specified  subset of a real line $\mathbb{R}$ (e.g. $[c, \infty)$).

  A point estimator for the inverse set can be constructed as $\hat{\mu}_n^{-1}(U)=\{s \in S: \hat \mu(s) \in U\}$, where $\hat{\mu}_n$ is an empirical estimator of $\mu$ based on $n$ observations. To quantify the spatial uncertainty of this estimate, we define the “lower bound” and “upper bound” as follows: 
  $\text{CR}_{\text{in}}(U) \subseteq {\mu}_n^{-1}(U) \subseteq \text{CR}_{\text{out}}(U)$
  which satisfy: 
  $\mathbb{P}\left(\text{CR}_{\text{in}}(U) \subseteq {\mu}_n^{-1}(U) \subseteq \text{CR}_{\text{out}}(U)\right) \geq 1 - \alpha$
  for a pre-specified confidence level $1-\alpha$ (e.g., $\alpha = 0.05$).  

  \subsection{Construction of Confidence Region (CR) for Inverse Set}
  
  Our construction of the simultaneous confidence regions for the inverse set follows the procedure in \cite{Ren:2024}. The framework applied in the package generalizes the estimation of such inverse sets to dense and non-dense domains with protection against inflated Type I error, and constructs multiple upper, lower or interval confidence regions of $\mu^{-1}(U)$ over arbitrarily chosen thresholds. The coverage probability is achieved non-asymptotically and simultaneously through inverting simultaneous confidence intervals. For instance, suppose we are interested in inverse set  $\mu^{-1}([c,\infty))$ for a single value $c$, the inverse confidence regions are constructed by inverting simultaneous confidence bands (SCBs). Given SCB bounds $\hat{B}_{l}(\boldsymbol{s})$ and $\hat{B}_{u}(\boldsymbol{s})$ satisfying \[\mathbb{P}\left(\forall\boldsymbol{s}\in\mathcal{S}: \hat{B}_{l}(\boldsymbol{s}) \leq \mu(\boldsymbol{s}) \leq \hat{B}_{u}(\boldsymbol{s})\right) = 1-\alpha.\]

  The inner and outer confidence regions for the inverse upper excursion set $\mu^{-1}[c, \infty)$ can be calculated as:  
  $$\text{CR}_{\text{in}}[c, \infty) := \hat{B}_\ell^{-1}[c, \infty)$$
  $$\text{CR}_{\text{out}}[c,\infty) := \hat{B}_u^{-1}[c, \infty)$$
  These CRs are valid for all $c \in \mathbb{R}$. Therefore we name them simultaneous confidence regions (SCRs).\\
  The outer and inner confidence regions for the inverse lower excursion set $\mu^{-1}\left(-\infty, c\right]$ are calculated as:
  $$\text{CR}_{\text{in}}\left(-\infty, c\right] := \hat{B}_u^{-1}\left(-\infty, c\right] = \left( \hat{B}_u^{-1}\left[c, +\infty\right) \right)^{\complement}$$
  $$\text{CR}_{\text{out}}\left(-\infty, c\right] := \hat{B}_\ell^{-1}\left(-\infty, c\right]
  = \left( \hat{B}_\ell^{-1}\left[c, +\infty\right) \right)^{\complement}$$
  
  The inner and outer confidence regions for the inverse interval set $\mu^{-1}[a, b]$ are calculated as:
  $$\text{CR}_{\text{in}}[a, b] := \hat{B}_\ell^{-1}[a, \infty) \cap \hat{B}_u^{-1}(-\infty, b]$$
  $$\text{CR}_{\text{out}}[a, b] := \hat{B}_u^{-1}[a, \infty) \cap \hat{B}_\ell^{-1}(-\infty, b]$$
  
  \subsection{Construction of SCB for Functional Regression Model}

  
The SCoRES package provides the construction of SCBs for the linear function-on-scalar regression (FoSR) model. 
A simple example is given by
\[
  Y_i(t) = \beta_0(t) + \beta_1(t) X_{i1} + b_i(t) + \epsilon_i(t),
\]
where $Y_i(t)$ is a functional outcome for the $i$th subject, $X_{i1}$ is a scalar covariate, each $\beta_j(t)$ is a coefficient function, $b_i(t)$ is a subject-specific functional random effect that captures correlation within subjects not explained by the mean, and $\epsilon_i(t)$ are normally distributed i.i.d. errors. A standard estimation framework \citep{crainiceanu2024functional} involves two steps. First, fit the mean model
\[
Y_i(t) = \beta_0(t) + \beta_1(t) X_{i1} + e_i(t).
\]
Then, estimate the residual process $e_i(t) = b_i(t) + \epsilon_i(t)$ using FPCA, where $b_i(t)$ follows a mean-zero Gaussian process with covariance $\Sigma$.  
If $\phi_k(\cdot)$ are eigenfunctions of the covariance operator $K_X$, then
\[
b_i(t) = \sum_{k=1}^{\infty} \xi_{ik} \phi_k(t).
\]
The resulting GAMM-FPCA model is
\[
Y_i(t) = \beta_0(t) + \beta_1(t) X_{i1} + \sum_{k=1}^{K} \xi_{ik} \phi_k(t) + \epsilon_i(t).
\]

We consider simultaneous confidence bands (SCBs) for a target function $\eta(\cdot)$ on a grid $\mathcal{S}$. For example, $\eta(\cdot)$ could be $Y_i(t)$ or $\beta_1(t)$.
Given an estimator $\hat{\eta}_N(s)$ with pointwise standard error $\hat{\zeta}_N(s)$ and a normalizing
factor $\tau_N$, we can define the simultaneous confidence band for $\eta(\cdot)$ as:
\[
\mathrm{SCB}(s;q_{\alpha,N})
= \Big[\, \hat{\eta}_N(s) - q_{\alpha,N}\tfrac{\hat{\zeta}_N(s)}{\tau_N},\;
              \hat{\eta}_N(s) + q_{\alpha,N}\tfrac{\hat{\zeta}_N(s)}{\tau_N} \Big].
\]
And we can verify that these bands achieve $(1-\alpha)$ simultaneous coverage by
\[
\mathbb{P}\!\left(\forall s\in\mathcal{S}:\; \eta(s)\in \mathrm{SCB}(s;q_{\alpha,N})\right)=1-\alpha,
\]
whenever the critical value $q_{\alpha,N}$ satisfies
\[
\mathbb{P}\!\left(\max_{s\in\mathcal{S}}\, \tau_N\,
\frac{\hat{\eta}_N(s)-\eta(s)}{\hat{\zeta}_N(s)} \;>\; q_{\alpha,N}\right)=\alpha. 
\]

The $q_{\alpha,N}$ is unknown in practice. In SCoRES, we implement two approaches for estimating the $q_{\alpha,N}$ for functional coefficient functions in FoSR model: a simulation based procedure from \cite{crainiceanu2024functional} denoted Correlation and Multiplicity-Adjusted (CMA) bands, and a multiplier bootstrap procedure. We detail the CMA procedure below:

\begin{enumerate}
  \item Simulate model parameters
    $\boldsymbol{\beta}_1, \ldots, \boldsymbol{\beta}_B \overset{\text{i.i.d.}}{\sim} \mathcal{N}(\hat{\boldsymbol{\beta}}, \hat{V}_{\boldsymbol{\beta}})$,
    where $(\hat{\boldsymbol{\beta}}, \hat{V}_{\boldsymbol{\beta}})$ are estimated from a fitted FoSR model.

  \item For each $b=1,\ldots,B$, compute
\[
   \mathbf{X}_b \;=\; \frac{\mathbf{B}(\beta_b - \hat{\beta})}{\mathbf{D}_f},
\]
    where the division is element-wise and $\mathbf{B}$ maps parameters to functional effects.

  \item Define
\[
   d_b \;=\; \max\!\big(|\mathbf{X}_b|\big), \quad b=1,\ldots,B,
\]
    where the absolute value is taken element-wise.

  \item Estimate $q_{\alpha,N}$ as the $(1 - \alpha) \cdot 100\%$ quantile of $\{d_1,\ldots,d_B\}$.
\end{enumerate}

The second is the multiplier-t bootstrap procedure for constructing confidence bands. A complete introduction to the method can be found in previous work \cite{Telschow:2022}.

\begin{enumerate}
  \item Compute residuals $R_1^N, \ldots, R_N^N$, where $R_n^N = \sqrt{\frac{N}{N - 1}} \left( Y_n - \hat{\mu}_N \right)$, and multipliers $g_1, \ldots, g_N$ \\ $\overset{\text{i.i.d.}}{\sim} g$ with $\mathbb{E}[g] = 0$ and $\mathrm{var}[g] = 1$, where $\hat{\mu}_N$ is fitted mean value from model (e.g. FoSR).

  \item Estimate $\hat{\epsilon}_N^*(s)$ from $g_1 Y_1(s), \ldots, g_N Y_N(s)$.

  \item Compute 
\[
T^*(s) = \frac{1}{\sqrt{N}} \sum_{n=1}^N g_n \frac{R_n^N(s)}{\hat{\epsilon}_N^*(s)}
\]

  \item Repeat steps 1 to 3 many times to approximate the condition law $\mathcal{L}^* = \mathcal{L}\!\left(T^* \mid Y_1, \ldots, Y_N\right)$. Take the $(1 - \alpha) \cdot 100\%$ quantile of $\mathcal{L}^*$ to estimate $q_{\alpha, N}$.
\end{enumerate}

At step 2, we allow three choices for the multiplier distribution $g_i$, each with mean zero and unit variance:
\begin{itemize}
  \item \texttt{rademacher}: $g_i \in \{-1,+1\}$ with equal probability;
  \item \texttt{gaussian}: $g_i \sim \mathcal{N}(0,1)$;
  \item \texttt{mammen}: a two–point distribution with mean $0$ and variance $1$ from \cite{Mammen:1993}.
\end{itemize}
Unless stated otherwise, we use the Rademacher multipliers.

At step 3, we consider two alternatives for the pointwise standard errors
$\hat{\epsilon}_N^*(s_j)$:
\begin{itemize}
  \item \emph{regular},
  \[
    \hat{\epsilon}_N^*(s_j)
    = \sqrt{\frac{1}{n}\sum_{i=1}^n\big(Y_i(s_j)-\hat{\beta}(s_j)\big)^2/(n-1)}\,;
  \]
  \item \emph{t},
  \[
    \hat{\epsilon}_N^*(s_j)
    = \sqrt{\frac{N}{N-1}\,\Big|\mathbb{E}_b\!\big[Y^{b}(s_j)^2\big]
      - \big(\mathbb{E}_b[Y^{b}(s_j)]\big)^2\Big|},
  \]
  where expectations are taken over bootstrap replicates and $Y^{b}(s_j)$ denotes the perturbed sample at iteration $b$. The absolute value improves numerical stability when subtracting nearly equal quantities.
\end{itemize}
Unless noted otherwise, we report results using the \emph{t} estimator.

\subsection{Construction of SCB for Linear/Logistic Regression Model}

We next describe the bootstrap algorithm used to construct simultaneous confidence intervals (SCIs) for the mean outcome of regression on a fixed test design matrix. We also use the same procedures for the construction of SCIs for the regression coefficients. For details of this algorithm, please refer to \cite{Ren:2024}.

Suppose we have training data outcome $y$, design matrix $X$, and a fixed test design matrix $\tilde{X}$. Let $f(\beta, X)$ denote the fitted regression function. The algorithm proceeds as follows.

First, we estimate $\hat{\beta}$ on the training data $(y, X)$ using least squares. Then we compute the estimated mean outcome on the test design matrix $\hat{E}(\tilde{y}) := f(\hat{\beta}, \tilde{X})$,
together with its standard deviation $\hat{\sigma}$.

For bootstrap samples $b=1,\ldots,L$, repeat:
\begin{enumerate}
  \item Resample $(y_b, X_b)$ with replacement from the training data.
  \item Fit the model on the resampled data to obtain $\hat{\beta}_b$.
  \item Compute the estimated mean $\hat{E}(\tilde{y}_b):=f(\hat{\beta}_b, \tilde{X})$ and its pointwise standard deviation $\hat{\sigma}_b$.
  \item Calculate the standardized absolute residuals on the test design grid,
  \[
  r_b = \frac{\big| \hat{E}(\tilde{y}_b) - \hat{E}(\tilde{y}) \big|}{\hat{\sigma}_b}.
  \]
  \item Record the maximum value of $r_b$ as $r^{\max}_b$.
\end{enumerate}

After $L$ bootstrap iterations, take the $(1-\alpha)$ quantile of the empirical distribution of $\{r^{\max}_b\}_{b=1}^L$ as the threshold $a$. The SCI on the test design matrix is then given by
\[
\Big( \hat{E}(\tilde{y}) - a \times \hat{\sigma}, \;\; \hat{E}(\tilde{y}) + a \times \hat{\sigma} \Big).
\]
For logistic regression, the SCI is further transformed back to the data scale using the link function.

\subsection{Construction of SCB for Spatial Generalized Least Square Model}

We also provide tools in SCoRES for estimating SCB of geographic data, however, its use is currently limited to the spatial generalized least squares model. 

Let the spatial domain be sampled on spots \(s\in\mathcal S\) with coordinates \(s=(x_s,y_s)\).
At each spot \(s\), we observe \(n\) outcomes \(\mathbf z_s\in\mathbb R^{n}\) with design matrix
\(\mathbf X\in\mathbb R^{n\times p}\). We fit a spot-specific generalized least squares (GLS) model
\begin{equation}
  \mathbf z_s=\mathbf X\boldsymbol\beta(s)+\boldsymbol\varepsilon_s,\qquad
  \boldsymbol\varepsilon_s\sim \mathcal N\!\left(\mathbf 0,\ \mathbf V_s\right),\notag
\end{equation}
where the variance-covariance matrix within the spot encodes the prespecified correlation structure $\mathbf V_s$.

SCoRES allows users to directly input the matrix $V_s$ for each spot $s$, or specify the correlation structure $\mathbf R$.

Assume we are interested in the linear functional $\eta(s)=\mathbf w^\top \boldsymbol\beta(s)$, with $\mathbf w$ as a weight vector for the specific linear combination. The pointwise estimated standard error is $\widehat{\zeta}(s)=\sqrt{\mathbf w^\top Var(\beta) \mathbf w}$

To quantify uncertainty uniformly across spots, we construct a simultaneous confidence band (SCB)
for \(\eta(\cdot)\) on the grid \(\mathcal S\):
\begin{equation}
  \mathrm{SCB}(s;\,q_\alpha)=\Big[\ \widehat{\eta}(s)-q_\alpha\,\widehat{\zeta}(s),\ \
  \widehat{\eta}(s)+q_\alpha\,\widehat{\zeta}(s)\ \Big],\qquad s\in\mathcal S,\notag
\end{equation}
where \(q_\alpha\) satisfies
\begin{equation}
  \mathbb P\!\left(\max_{s\in\mathcal S}\frac{|\widehat{\eta}(s)-\eta(s)|}{\widehat{\zeta}(s)}
  \le q_\alpha\right)= 1-\alpha.\notag
\end{equation}

We estimate $q_\alpha$ by multiplier-t bootstrap method introduced before.

  \subsection{Simulation Studies}

We conduct a series of simulation studies to evaluate the coverage probabilities of the simultaneous confidence bands (SCBs) constructed using the SCoRES package for linear, logistic, and functional outcomes.

To assess the validity of the CMA and multiplier-t bootstrap methods for constructing SCBs of functional parameters, we simulate data from the following model, fit a GAMM-FPCA model, and compute the SCB for $\hat \beta_1(t)$ using the \texttt{SCB\_functional\_outcome()} function in our package.

\begin{align*}
&t_1, t_2, \ldots, t_{50} \text{ are equally spaced points in } [0, 1],\\
&\beta_0 = 0, \quad \beta_1(t) = \sin(6\pi t), \\
&x_i \sim \text{Bernoulli}(0.6), \\
&b_i(t) = \sum_{k=1}^5 \xi_{ik} \phi_k(t), \quad \xi_{ik} \sim \mathcal{N}(0, \sigma_k^2), \\
&\epsilon_{it} \sim \mathcal{N}(0, 0.25), \\
&Y_{it} = \beta_0 + \beta_1(t) x_i + b_i(t) + \epsilon_{it}.
\end{align*}

To examine the performance of the nonparametric bootstrap method for constructing SCBs of conditional mean functions in linear regression, we generate data according to the following model and compute the SCB for $E(y|x_1)$ using \texttt{SCB\_linear\_outcome()}:

\begin{align*}
x_1 &\sim \mathcal{N}(0, 1), \\
\epsilon &\sim \mathcal{N}(0, \sqrt{2}), \\
y &= -1 + x_1 + 0.5 x_1^2 - 1.1 x_1^3 + \epsilon.
\end{align*}

Similarly, for the logistic regression setting, we simulate data from the model below and compute the SCB for $E(y|x_1)$ using \texttt{SCB\_logistic\_outcome()}:

\begin{align*}
x_1 &\sim \mathcal{N}(0, 1), \\
\mu(x_1) &= -1 + x_1 + 0.5 x_1^2 - 1.1 x_1^3, \\
p(x_1) &= \frac{1}{1 + e^{-\mu(x_1)}}, \\
y &\sim \text{Bernoulli}(p(x_1)).
\end{align*}

Finally, we assess the coverage performance of SCBs for regression coefficients using the \\
\texttt{SCB\_regression\_coef()} function. We simulate both linear and logistic regression models with multivariate covariates $\beta_i, i=1,\cdots,M,$ where $M=5$.

\textbf{Linear model:}
\begin{align*}
X &\sim \mathcal{N}(0, \Sigma), \quad \Sigma_{ij} = \rho^{|i - j|}, \quad \rho = 0.4, \\
\beta &\sim \mathcal{N}(0, I_M), \\
\epsilon &\sim \mathcal{N}(0, 1), \\
y &= X \beta + \epsilon.
\end{align*}

\textbf{Logistic model:}
\begin{align*}
X &\sim \mathcal{N}(0, \Sigma), \quad \Sigma_{ij} = \rho^{|i - j|}, \quad \rho = 0.4, \\
\beta &\sim \mathcal{N}(0, I_M), \\
\mu &= X \beta, \\
p &= \frac{1}{1 + e^{-\mu}}, \\
y &\sim \text{Bernoulli}(p).
\end{align*}

\begin{figure}[H]
  \centering
    \includegraphics[height=5.5cm, width=\textwidth]{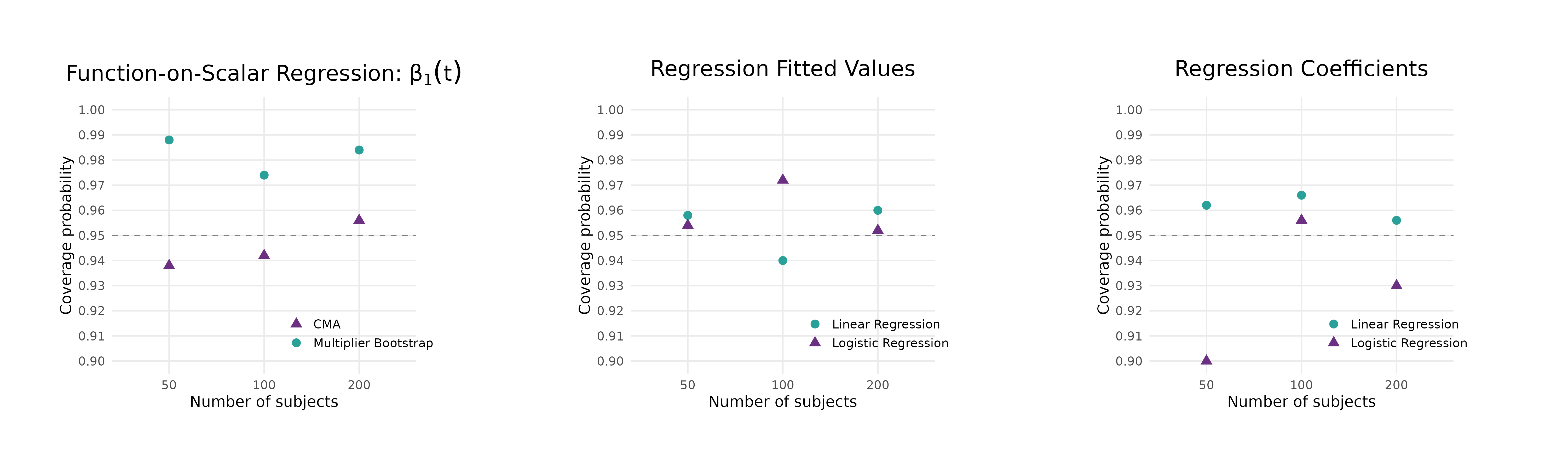}
  \caption{
\textbf{Performance of SCB coverage probabilities in simulated datasets under multiple datasets. } Left panel: Coverage probabilities for
the coefficient function $\beta_1(t)$ from a function-on-scalar regression
(FoSR) model, comparing CMA and the multiplier-t bootstrap across sample sizes
of 50, 100, and 200. Middle panel: Coverage probabilities for
regression fitted values obtained from linear and logistic regression models. Right panel: Coverage probabilities for regression
coefficient estimates from linear and logistic regression models across the same
sample sizes. Dashed horizontal lines mark the nominal 95\% coverage level. Coverage probabilities were computed from 500 independent simulation replicates.}
  \label{fig:simulation}
\end{figure}

Each simulation setting is repeated 500 times to empirically evaluate the coverage performance of the constructed SCBs. Figure~\ref{fig:simulation} displays the empirical coverage probabilities of the SCBs for the coefficient functions from a FosR model, regression outcomes and coefficients from linear/logistic models, shown separately for each setting.

\section{The SCoRES Package}
\label{sec:software}
\noindent
\textbf{SCoRES} is available to the public via 
\href{https://cran.r-project.org}{The Comprehensive R Archive Network (CRAN)}. 
To download, one can use the following code:

\begin{verbatim}
R> install.packages("SCoRES")
\end{verbatim}

\noindent 
The development version is available on 
\href{https://github.com/AngelaYuStat/SCoRES}{GitHub}. 
This can be downloaded using the \textbf{devtools} package with the following command 
\citep{Wickham:2025}:

\begin{verbatim}
R> devtools::install_github("AngelaYuStat/SCoRES")
\end{verbatim}

  \subsection{Operation}

Users should have R version 4.4 or higher to use \textbf{SCoRES}. 
The \textbf{SCoRES} package is designed to estimate the simultaneous confidence region for inverse sets, and construct SCBs for multiple regression models.
The package functions fall into three categories: SCB estimation, confidence region estimation, and plotting.
There are also three datasets provided.

  \subsection{Functions}
  The main functions of the \textbf{SCoRES} package are
  \texttt{SCB\_linear\_outcome()}, \\
  \texttt{SCB\_logistic\_outcome()}, 
  \texttt{SCB\_regression\_coef()},
  \texttt{SCB\_functional\_outcome()} and \\
  \texttt{SCB\_gls\_geospatial()}, to obtain estimates of SCBs for target functions, \texttt{scb\_to\_cs()} for converting the input SCBs to simultaneous confidence regions, and 
  \texttt{plot\_cs()} for visualizing both SCBs and the simultaneous confidence region. 
  \texttt{SCB\_linear\_outcome()}, \\ \texttt{SCB\_logistic\_outcome()} and \texttt{SCB\_regression\_coef()} require data and a model formula as input. They first fit the corresponding models and then use a nonparametric bootstrap algorithm to construct SCBs for mean outcome functions and SCIs for coefficients.

  \subsubsection{SCB Related Functions}
  For \texttt{SCB\_linear\_outcome()} and \texttt{SCB\_logistic\_outcome()}, several arguments allow the user to tailor the procedure. The \texttt{df\_fit} argument supplies a data frame containing the training design matrix used to fit the regression model, while \texttt{grid\_df} specifies the test/design grid at which the mean outcome is evaluated and the simultaneous confidence band (SCB) is constructed. The model formula is provided via \texttt{model} (e.g., \texttt{y \textasciitilde\ x1 + x2}). For logistic regression, estimates and bands are returned on the response (probability) scale after applying the inverse link function transformation.

  Additional arguments control the bootstrap step. The \texttt{n\_boot} argument sets the number of bootstrap samples used to generate the empirical distribution for estimating $q_\alpha$ (default \texttt{1000}). The \texttt{alpha} argument specifies the significance level for the band (e.g., \texttt{0.05} for a 95\% band). The default value is \texttt{0.05}. An optional \texttt{grid\_df\_boot} argument (available for \texttt{SCB\_linear\_outcome()} only) can be used to provide a separate evaluation grid during the bootstrap step. If omitted, \texttt{grid\_df} is used.

  \texttt{SCB\_regression\_coef()} computes simultaneous confidence bands (SCBs) for the model coefficients using a nonparametric bootstrap procedure. By specifying the \texttt{type} argument, users can choose to fit either a linear or logistic regression model. The default option is linear regression.

  The input arguments are mainly consistent with those of \texttt{SCB\_linear\_outcome()} and \\
  \texttt{SCB\_logistic\_outcome()}, with the key difference that no testing design matrix (\texttt{grid\_df}) is required.
  
  \texttt{SCB\_functional\_outcome()} accepts a FoSR functional regression object as input, and allows users to specify the target function format via arguments. It obtains point estimates through \texttt{mean\_response\_predict()} and supports two approaches for SCB construction: (i) the parametric CMA method (\texttt{cma()}) and (ii) a multiplier bootstrap implemented internally via \texttt{SCB\_dense} and \texttt{MultiplierBootstrap}.

  The function takes a functional data frame \texttt{data\_df} that contains the variable names and values for the functional outcome, its domain (e.g., time), and the ID (e.g., subject) used to fit the model. The \texttt{object} argument takes a fitted function-on-scalar regression (FoSR) object (e.g., from \texttt{mgcv::gam()} or \texttt{mgcv::bam()}). The \texttt{method} argument selects the SCB construction approach:
  \texttt{"cma"} (Correlation and Multiplicity Adjusted) or
  \texttt{"multiplier"} (multiplier bootstrap). For \texttt{method = "multiplier"}, the outcome in \texttt{data\_df} must not be identically zero within any specified domain segment (except at domain index zero), otherwise an error is returned. If the outcome contains missing values (\texttt{NA}), they are imputed via \texttt{fpca.face} before performing the multiplier bootstrap.

  The \texttt{fitted} argument chooses the target: \texttt{TRUE} estimates SCBs for the fitted mean outcome function, and \texttt{FALSE} estimates SCBs for a fitted coefficient function. 
  The default value is \texttt{TRUE}. The band level is set by
  \texttt{alpha} (default \texttt{0.05}).

  Model-mapping arguments specify variables in \texttt{data\_df}: \texttt{outcome} names the functional outcome; \texttt{domain} names the domain variable (e.g., time); \texttt{id} names the subject ID. The \texttt{subset} argument is an atomic character vector (e.g., \texttt{c("use = 1", "age = 30")}) that specifies the target function for SCB construction. Each element must be of the form \texttt{<var> = <value>} for a scalar grouping variable, ignoring whitespace. Binary or categorical character variables should be converted to numeric in advance. Factors are not allowed because factor inputs are expanded to indicator variables in the design matrix, potentially changing variable names. If \texttt{subset = NULL}, the reference group is used.

  Bootstrap controls include \texttt{nboot} (number of bootstrap samples; default \texttt{10000} for \texttt{"cma"}, \texttt{5000} for \texttt{"multiplier"}), \texttt{method\_SD} (the pointwise SD estimator, \texttt{"t"} or \texttt{"regular"}; default \texttt{"t"}), and \texttt{weights} (multiplier type: \texttt{"rademacher"}, \texttt{"gaussian"}, or \texttt{"mammen"}; default \texttt{"rademacher"}).
  
  \texttt{SCB\_gls\_geospatial()} takes 2D geospatial observations and a design matrix as inputs. It then fits the corresponding GLS model and employs a multiplier bootstrap algorithm to construct the SCBs for specified linear combinations of coefficients functions. 

  The input \texttt{sp\_list} contains the spatial coordinates
  (\texttt{x}, \texttt{y}) and a 3D array of repeated observations at each spot. A design matrix is supplied via \texttt{data\_fit}, with rows corresponding to observations and columns to covariates (the first column typically an intercept). Categorical variables should be converted to dummy variables in advance.  

  The argument \texttt{w} specifies a linear combination of GLS coefficients for which the SCBs are constructed. Correlation within spots can be modeled by setting \texttt{correlation} (e.g., \texttt{"corAR1"}, \texttt{"corCompSymm"}) with optional parameters in \texttt{corpar}, or by providing group identifiers in \texttt{groups}. Alternatively, a per–spot covariance array \texttt{V} can be supplied directly. If neither correlation nor \texttt{V} is provided, the procedure reduces to ordinary
  least squares (OLS).  

  Construction of SCB is performed using a multiplier bootstrap with \texttt{N} bootstrap samples (default \texttt{1000}). The significance level is controlled by \texttt{alpha} (default \texttt{0.1}). An optional logical matrix \texttt{mask} can be
  used to exclude locations (e.g., water areas) from SCB computation.  

  For all the above SCB related functions, the output is a list containing matrices of upper and lower SCB bounds (\texttt{scb\_up}, \texttt{scb\_low}), estimated means (\texttt{mu\_hat}), and other additional metadata relevant to the model fitting procedure.

\subsubsection{Confidence Region Related Functions}

  Given simultaneous confidence bands (SCBs) on a one– or two–dimensional domain, \texttt{scb\_to\_cs()} converts these bands into simultaneous confidence regions for inverse sets of upper excursion sets, lower excursion sets, and interval sets. Users can either use our SCB-related functions to construct SCBs for input, or they can directly input their own SCBs based on any models to construct SCRs.
  
  The function takes the upper and lower band limits: \texttt{scb\_up} and \texttt{scb\_low} (1D vectors or 2D matrices), and a threshold specification 
  \texttt{levels}. When \texttt{type} is \texttt{"upper"}, \texttt{"lower"}, or \texttt{"two-sided"}, \texttt{levels} is a numeric vector of thresholds. For interval sets (\texttt{type = "interval"}), \texttt{levels} is a list with named elements 
  \texttt{low} and \texttt{up} defining \([\,\texttt{low},\texttt{up}\,]\). For the \texttt{"two-sided"} option, the procedure estimates outer CSs for both upper and 
  lower excursions. Optional arguments \texttt{x1} and \texttt{x2} supply plotting coordinates for the first and second dimensions, respectively. 

  The function returns a compact list that records (i) the thresholds used (\texttt{levels}); (ii) the inner and outer logical maps for each level (\texttt{U\_in}, \texttt{U\_out}; or, for \texttt{"two-sided"}, lower/upper containment maps); (iii) optional containment summaries (\texttt{contain\_individual}, \texttt{contain\_all}) when \texttt{true\_mean} is supplied; and 
  (iv) optional \texttt{ggplot2} objects (\texttt{plot\_cs}) when visualization is requested.

\begin{figure}[H]
\centering
\includegraphics[width=.8\linewidth]{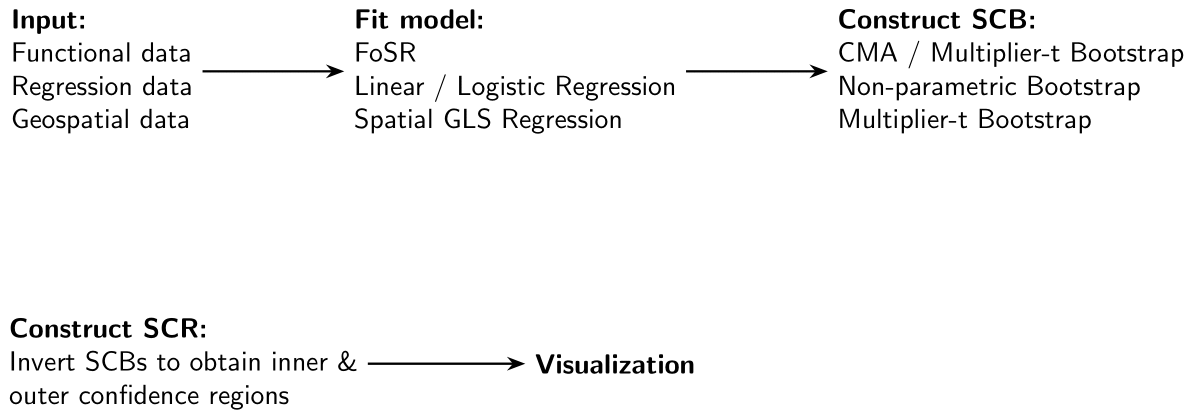}
\caption{Overall workflow of SCoRES}
\label{fig:SCoRES_workflow}
\end{figure}

  The overall workflow of the \texttt{SCoRES} package is illustrated in Figure~\ref{fig:SCoRES_workflow}. The figure shows the general procedure for constructing SCBs and SCRs. Starting from functional, regression, or geospatial data, users can fit models such as functional-on-scalar regression (FoSR), linear or logistic regression, and spatial GLS regression. SCBs can then be constructed using CMA, non-parametric bootstrap, or multiplier-\(t\) bootstrap methods. Finally, simultaneous confidence regions are obtained by inverting the SCBs to form inner and outer confidence envelopes, which can be visualized to highlight regions where the estimated effects exceed a meaningful threshold.

\begin{figure}[H]
\centering
\includegraphics[width=.8\linewidth]{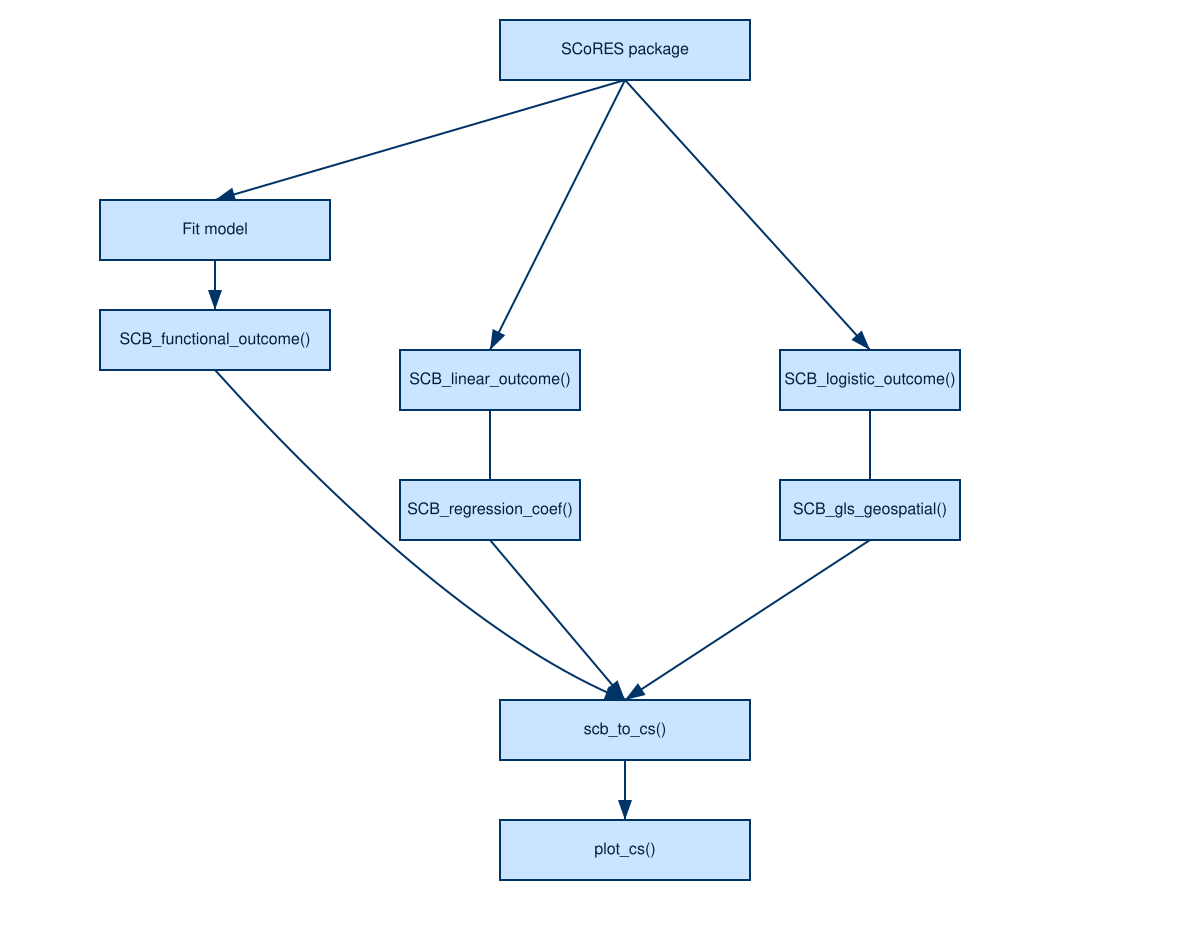}
\caption{Function structure of SCoRES}
\label{fig:SCoRES_functions}
\end{figure}

Figure~\ref{fig:SCoRES_functions} presents the function structure within the package. 
It demonstrates how the core functions (\texttt{SCB\_functional\_outcome()}, \texttt{SCB\_linear\_outcome()}, 
\texttt{SCB\_logistic\_outcome()}, \texttt{SCB\_gls\_geospatial()}, and \texttt{SCB\_regression\_coef()}) connect 
to the post-processing and visualization modules (\texttt{scb\_to\_cs()} and \texttt{plot\_cs()}), 
providing a unified and extensible workflow from model fitting to simultaneous inference.\\
Because the modeling forms for functional data vary widely (e.g., different basis expansions, smoothness penalties, or estimation frameworks), 
\texttt{SCoRES} does not provide a general-purpose function for FoSR model fitting. 
Instead, users are encouraged to fit their own FoSR models—using packages such as \texttt{refund} or \texttt{mgcv}—and then supply the fitted model 
object as the input to the \texttt{SCB\_functional\_outcome()} function for constructing SCBs based on the estimated coefficient functions. 
This design provides maximum flexibility and ensures compatibility with diverse functional regression frameworks.\\
Moreover, users can also directly input their own SCBs obtained from any arbitrary model. The package can automatically construct the corresponding SCRs, and visualize the results using the same unified \texttt{scb\_to\_cs()} and \texttt{plot\_cs()} functions. 
This flexibility allows \texttt{SCoRES} to serve as a general framework for visualizing uncertainty regions across a wide range of statistical models.

  \subsection{Other Package Elements}
  \texttt{plot\_cs()} visualizes the inversion of simultaneous confidence bands (SCBs) into simultaneous confidence regions for upper or lower excursion sets on discrete, one–dimensional, or two–dimensional domains. It supports multi–level displays and annotated contours, using band plots in 1D and contour plots in 2D.

  The function expects an object \texttt{SCB} containing upper and lower bands. Thresholds are provided via \texttt{levels}. The \texttt{type} argument specifies whether to visualize upper or lower excursion sets (\texttt{"upper"} by default). Axes are defined by \texttt{x} and optional \texttt{y}. A mean function should be supplied: either the estimated mean \texttt{mu\_hat} or the true mean \texttt{mu\_true}. Plotting options include \texttt{together} (draw all levels on a single panel or one plot per level), axis labels \texttt{xlab}/\texttt{ylab}, contour labeling controls \texttt{level\_label} and \texttt{min.size}, as well as palette and label colors via \texttt{palette} and \texttt{color\_level\_label}.

  The function returns a \texttt{ggplot2} object that overlays the SCBs with the corresponding simultaneous confidence regions for the specified thresholds, producing band plots in 1D or contour maps in 2D.

  Lastly, the SCoRES package contains three datasets.

  \subsection{Important Dependencies}
  The package imports a set of widely used R libraries to support data manipulation, modeling, and visualization. 
  For data wrangling and reshaping we rely on \pkg{dplyr}, \pkg{tidyr}, \pkg{forcats}, \pkg{tibble}, \pkg{reshape}, and \pkg{magrittr}, with \pkg{rlang} enabling non-standard evaluation. 
  Model fitting and numerical routines draw on \pkg{stats}, \pkg{MASS}, \pkg{Matrix}, \pkg{matrixStats}, \pkg{nlme}, and \pkg{refund}, covering linear models, generalized least squares, and function-on-scalar regression. 
  Graphics are produced with \pkg{ggplot2}, combined via \pkg{patchwork}, and enhanced for geospatial/contour displays using \pkg{metR}; low-level devices and color management use \pkg{grDevices}. 
  Additional utilities are provided by \pkg{utils}. 
  Together, these dependencies enable efficient data processing, robust estimation, and publication-ready figures.

\section{Illustrations}
\label{sec:examples}
To demonstrate the flexibility of the SCoRES package, we analyze a few example datasets for several different model types.

  \subsection{SCoRES on Functional-on-Scalar Regression Model}
  We first look at the application on functional-on-scalar regression models. After installing the package from CRAN or
 GitHub, we load the SCoRES library.
 \begin{verbatim}
 R> library(SCoRES)
 \end{verbatim}
 We will use the pupil dataset in the package. The dataset contains functional observations of pupil size percent change after a light stimulus collected on the right eye. Covariates include a binary indicator of cannabis use, age, gender, BMI, and alcohol for 127 subjects.

The function-on-scalar regression model we will fit is

\[
\text{percent\_change}_i(t) = \beta_0(t) + \beta_1(t) \text{use}_i + b_i(t) + \epsilon_i(t),
\]

where $\text{percent\_change}_i(t)$ is percent change in pupil size $t$ seconds after the light stimulus, and $\text{use}_i$ is a binary variable where 1 indicates recent cannabis use and 0 indicates no recent cannabis use. Domain variable $t$ is \texttt{seconds}. And the covariate is \texttt{use}, which is a binary variable (1:use, 0:no use).
  
We construct simultaneous outer and inner confidence regions for the domain sets. These regions are derived directly from the simultaneous confidence bands (SCB) obtained via Function-on-Scalar Regression (FoSR).

 We load the mgcv library, a useful package for fitting FoSR models, and the pupil data.

\begin{verbatim}
R> library(mgcv)
R> data(pupil)
\end{verbatim}

Before calculating the SCBs, we first process pupil data by fitting a mean GAM model, extracting residuals and performing FPCA using
\code{SCoRES::prepare_pupil_fpca()}, the function will return an
enhanced dataset including the FPCA-derived basis scores (Phi1, Phi2,
Phi3, Phi4) for Function-on-Scalar Regression (FoSR) analysis.

Following the FPCA-based data augmentation, we fit a FoSR model using \code{mgcv::bam()}, which allows efficient estimation of
Generalized Additive Mixed Models (GAMMs). The model formula is designed to capture both population-level smooth trends and subject-specific functional variation.

\begin{verbatim}
R> pupil_fpca <- SCoRES::prepare_pupil_fpca(pupil)
R> fosr_mod <- mgcv::bam(
  percent_change ~ s(seconds, k=30, bs="cr") +
                   s(seconds, by = use, k=30, bs = "cr") +
                   s(id, by = Phi1, bs="re") +
                   s(id, by = Phi2, bs="re") +
                   s(id, by = Phi3, bs="re") +
                   s(id, by = Phi4, bs="re"),
  method = "fREML", data = pupil_fpca, discrete = TRUE)
\end{verbatim}

After obtaining the FoSR object \code{fosr_mod}, simultaneous
confidence bands (SCB) can be constructed through
\code{SCoRES::SCB_functional_outcome()}.
The SCoRES package provides two options for calculating simultaneous
confidence bands (SCBs), specified via the \code{method} argument:
\code{cma}: Correlation and Multiplicity Adjusted (CMA) confidence
bands via parametric approach; and \code{multiplier}: Dense confidence bands via Multiplier-t Bootstrap method.

Here, we estimated SCBs using both options separately for the mean
outcome $\text{percent\_change}_i(t)$ of user group:
\[
E[\text{percent\_change}(t) | \text{use} = 1] = \beta_0(t) + \beta_1(t),
\]
where \(Y(t)\) is the functional outcome (percent\_change), and
\(X_1\) is a scalar covariate (use).

We analyze the user group by specifying
\code{subset = c("use = 1")}, and we set \code{fitted = TRUE} to construct the SCB for the fitted mean outcome function. If \code{fitted = FALSE}, it will construct the SCB for the fitted parameter function. By setting \code{method = "cma"}, we apply Correlation and Multiplicity Adjusted method for computing the $q_{\alpha, N}$.

\begin{verbatim}
# CMA approach
R> results_pupil_cma <- SCoRES::SCB_functional_outcome(
  data_df = pupil,
  object = fosr_mod,
  method = "cma",
  fitted = TRUE,
  alpha = 0.05,
  outcome = "percent_change",
  domain = "seconds",
  subset = c("use = 1"),
  id = "id")
  
R> results_pupil_cma <- tibble::as_tibble(results_pupil_cma)
R> plot_cs(results_pupil_cma,
        levels = c(-18, -20, -22, -24),
        x = results_pupil_cma$domain,
        mu_hat = results_pupil_cma$mu_hat,
        xlab = "Seconds",
        ylab = "Percent_Outcome",
        level_label = TRUE,
        min.size = 40,
        palette = "Spectral",
        color_level_label = "black")
\end{verbatim}

\begin{figure}[H]
\centering
\includegraphics[width=.8\linewidth]{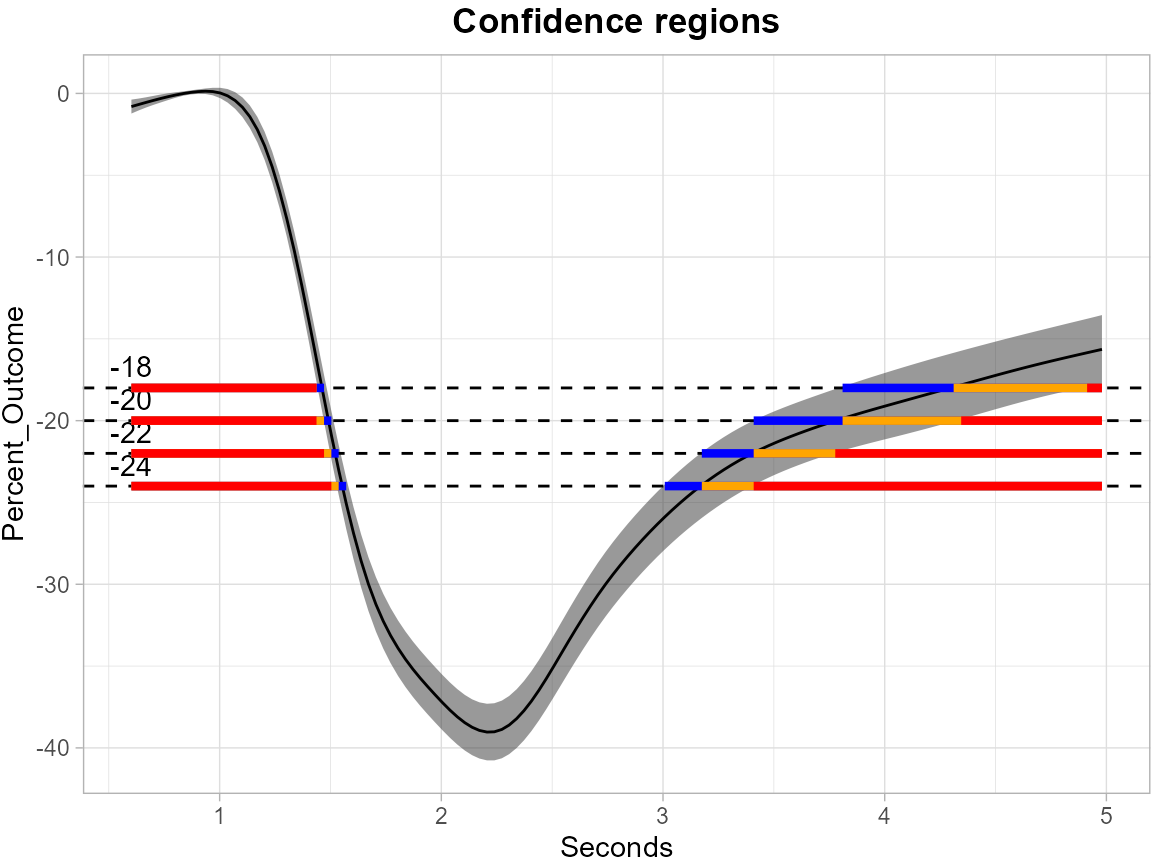}
\end{figure}

The plot demonstrates how to use SCB to find regions of \(s\)
where the estimated mean is greater than or equal to the four levels
-18, -20, -22 and -24. The gray shaded area is the 95\% SCB, the solid
black line is the estimated mean. The red horizontal line shows the
inner confidence region (where the lower SCB is greater than the
corresponding level) that is contained in the estimated inverse region;
the outer confidence region corresponds to where the upper SCB exceeds
the level.

By setting \code{method = "multiplier"}, we apply Multiplier-t Bootstrap method for computing the $q_{\alpha, N}$.

\begin{verbatim}
# Multiplier-t Bootstrap
R> results_pupil_multiplier <- SCoRES::SCB_functional_outcome(
  data_df = pupil,
  object = fosr_mod,
  method = "multiplier",
  fitted = TRUE,
  alpha = 0.05,
  outcome = "percent_change",
  domain = "seconds",
  subset = c("use = 1"),
  id = "id")
  
R> results_pupil_multiplier <- tibble::as_tibble(results_pupil_multiplier)
R> plot_cs(results_pupil_multiplier,
        levels = c(-18, -20, -22, -24),
        x = results_pupil_multiplier$domain,
        mu_hat = results_pupil_multiplier$mu_hat,
        xlab = "Seconds",
        ylab = "Percent_Outcome",
        level_label = TRUE,
        min.size = 40,
        palette = "Spectral",
        color_level_label = "black")
\end{verbatim}

\begin{figure}[H]
\centering
\includegraphics[width=.8\linewidth]{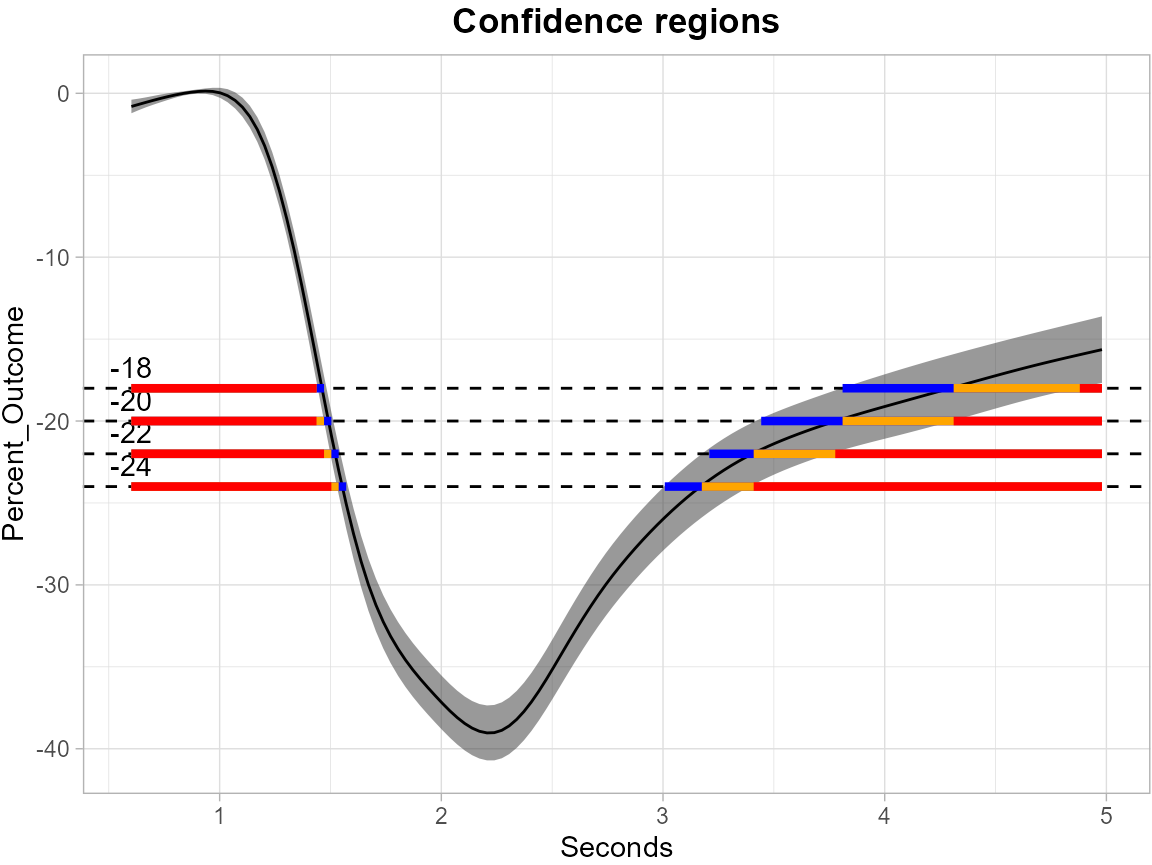}
\end{figure}

Next, we estimate SCBs using both options separately for the coefficient function for the user group $\beta_1(t)$.

\begin{verbatim}
# CMA approach for parameter function
R> results_pupil_cma_para <- SCoRES::SCB_functional_outcome(
  data_df = pupil,
  object = fosr_mod,
  method = "cma",
  fitted = FALSE,
  alpha = 0.05,
  outcome = "percent_change",
  domain = "seconds",
  subset = c("use = 1"),
  id = "id")
  
R> results_pupil_cma_para <-tibble::as_tibble(results_pupil_cma_para)
R> plot_cs(results_pupil_cma_para,
        levels = c(4.5, 5, 5.5, 6),
        x = results_pupil_cma_para$domain,
        mu_hat = results_pupil_cma_para$mu_hat,
        xlab = "Seconds",
        ylab = "Percent_Outcome",
        level_label = TRUE,
        min.size = 40,
        palette = "Spectral",
        color_level_label = "black")
\end{verbatim}

\begin{figure}[H]
\centering
\includegraphics[width=.8\linewidth]{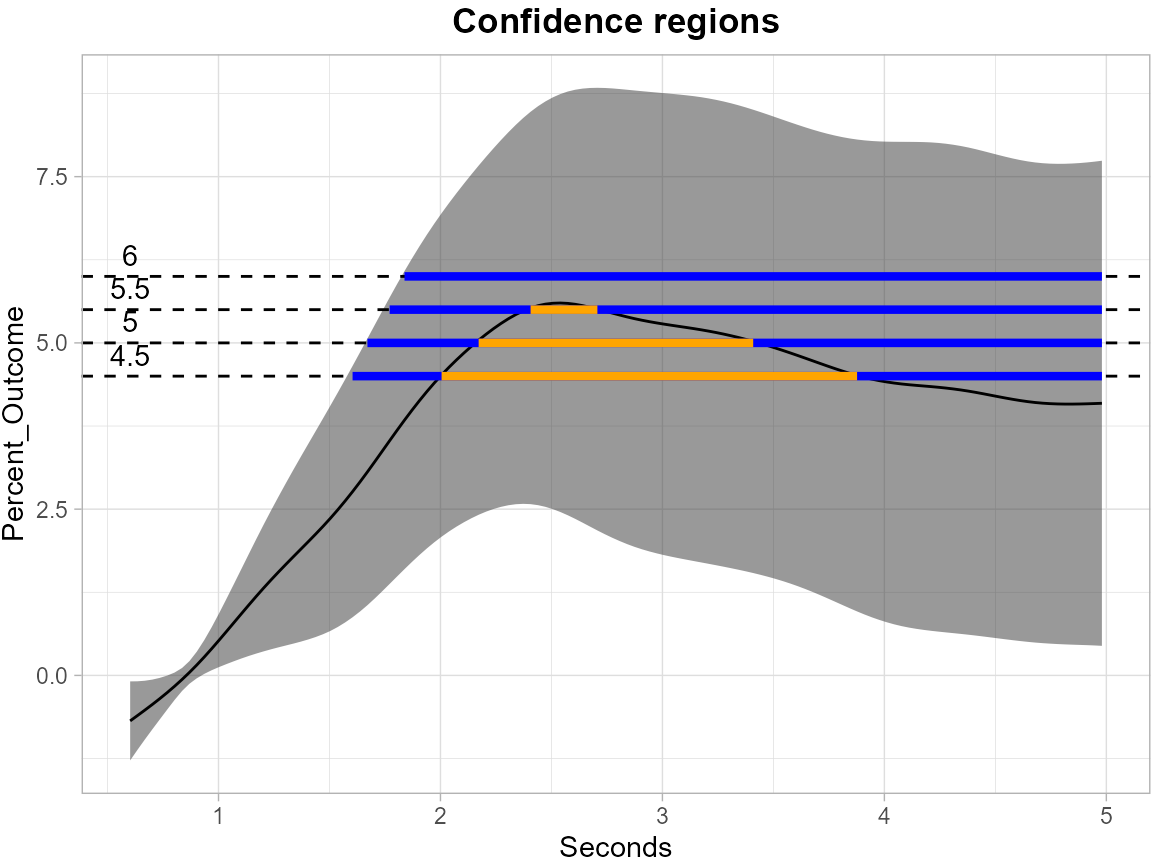}
\end{figure}

\begin{verbatim}
R> results_pupil_multiplier_para <- SCoRES::SCB_functional_outcome(
  data_df = pupil,
  object = fosr_mod,
  method = "multiplier",
  fitted = FALSE,
  alpha = 0.05,
  outcome = "percent_change",
  domain = "seconds",
  subset = c("use = 1"),
  id = "id")
  
R> results_pupil_multiplier_para <- 
   tibble::as_tibble(results_pupil_multiplier_para)
R> plot_cs(results_pupil_multiplier_para,
        levels = c(4.5, 5, 5.5, 6),
        x = results_pupil_multiplier_para$domain,
        mu_hat = results_pupil_multiplier_para$mu_hat,
        xlab = "Seconds",
        ylab = "Percent_Outcome",
        level_label = TRUE,
        min.size = 40,
        palette = "Spectral",
        color_level_label = "black")
\end{verbatim}

\begin{figure}[H]
\centering
\includegraphics[width=.8\linewidth]{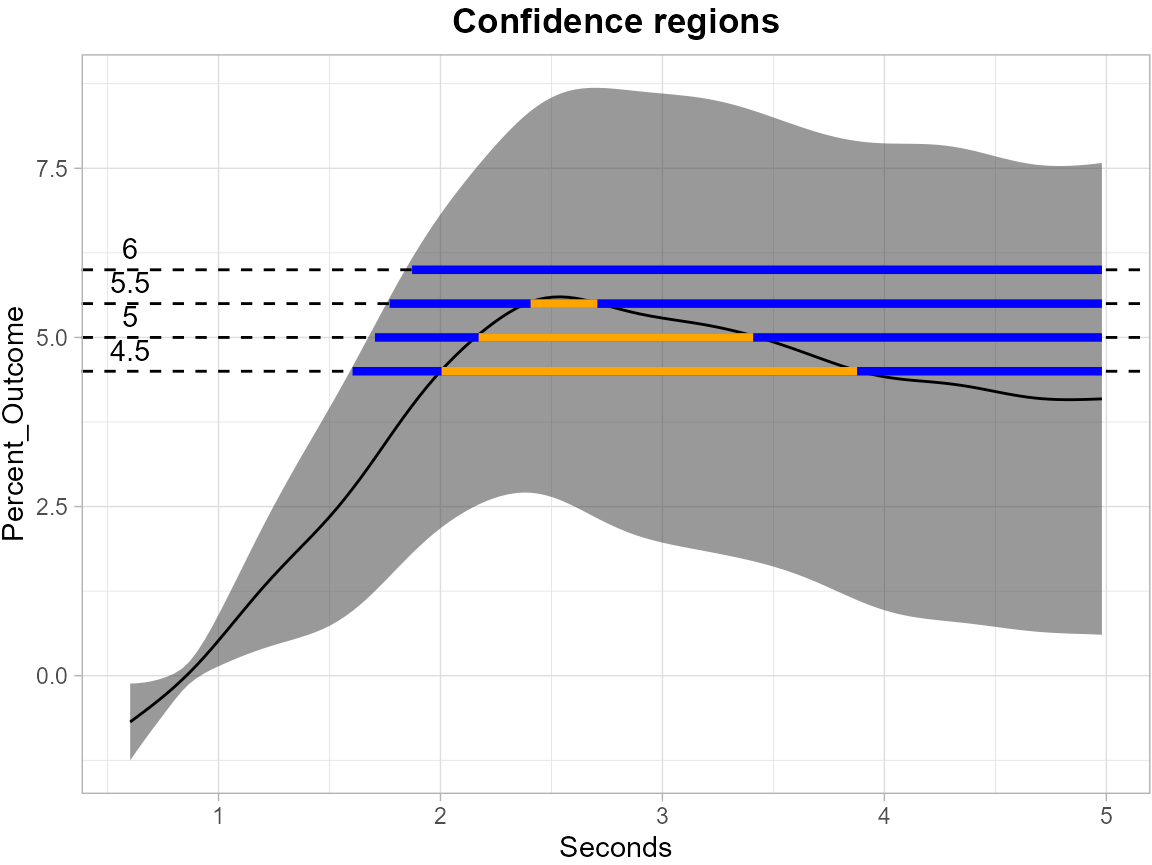}
\end{figure}

To further illustrate the power of SCoRES::SCB\_functional\_outcome for constructing SCBs for multiple group variables, we add age and gender as covariates, and analyze the 40-year-old male user group by specifying
`group\_name = c("use", "age", "gender")` and `group\_value = c(1, 40, 0)`. We set `fitted = TRUE`.

The function-on-scalar regression model is

\[
\text{percent\_change}_i(t) = \beta_0(t) + \beta_1(t) \text{use}_i + \beta_2(t) \text{age}_i + \beta_3(t) \text{gender}_i + b_i(t) + \epsilon_i(t),
\]

\begin{verbatim}
R> pupil_fpca <- SCoRES::prepare_pupil_fpca(pupil, example = "extended")
R> fosr_mod <- mgcv::bam(percent_change ~ s(seconds, k=30, bs="cr") +
            s(seconds, by = use, k=30, bs = "cr") +
            s(seconds, by = age, k = 30, bs = "cr") +
            s(seconds, by = gender, k = 30, bs = "cr") +
            s(id, by = Phi1, bs="re") +
            s(id, by = Phi2, bs="re") +
            s(id, by = Phi3, bs="re") +
            s(id, by = Phi4, bs="re"),
            method = "fREML", data = pupil_fpca, discrete = TRUE)
\end{verbatim}

\begin{verbatim}
# CMA approach
R> results_pupil_cma <- SCoRES::SCB_functional_outcome(
  data_df = pupil,
  object = fosr_mod,
  method = "cma",
  fitted = TRUE,
  alpha = 0.05,
  outcome = "percent_change",
  domain = "seconds",
  subset = c("use = 1", "age = 40", "gender = 0"),
  id = "id")

R> results_pupil_cma <- tibble::as_tibble(results_pupil_cma)
R> plot_cs(results_pupil_cma,
        levels = c(-18,-20,-22,-24),
        x = results_pupil_cma$domain,
        mu_hat = results_pupil_cma$mu_hat,
        xlab = "Seconds",
        ylab = "Percent_Outcome",
        level_label = TRUE,
        min.size = 40,
        palette = "Spectral",
        color_level_label = "black")
\end{verbatim}

\begin{figure}[H]
\centering
\includegraphics[width=.8\linewidth]{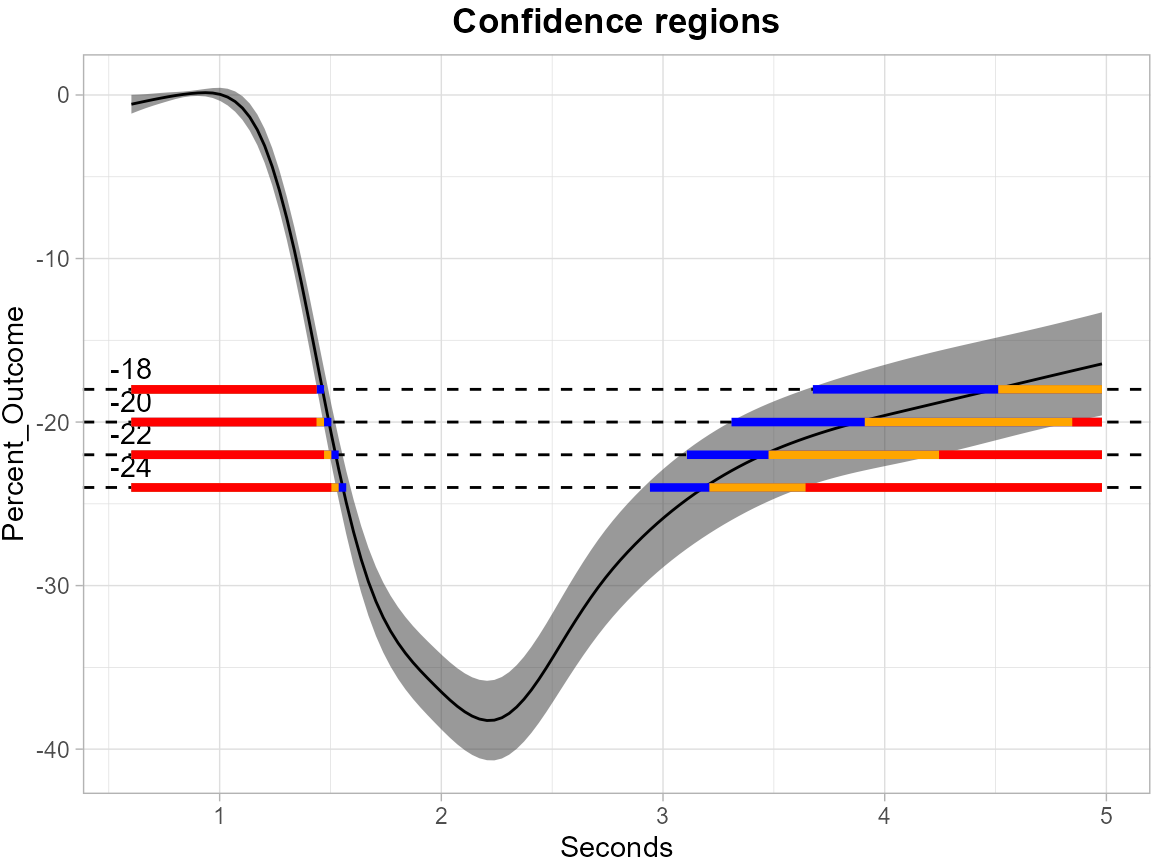}
\end{figure}

\begin{verbatim}
# Multiplier-t Bootstrap
R> results_pupil_multiplier <- SCoRES::SCB_functional_outcome(
  data_df = pupil,
  object = fosr_mod,
  fitted = TRUE,
  method = "multiplier",
  alpha = 0.05,
  outcome = "percent_change",
  domain = "seconds",
  subset = c("use = 1", "age = 40", "gender = 0"),
  id = "id")
  
R> results_pupil_multiplier <- tibble::as_tibble(results_pupil_multiplier)
R> plot_cs(results_pupil_multiplier,
        levels = c(-18,-20,-22,-24),
        x = results_pupil_multiplier$domain,
        mu_hat = results_pupil_multiplier$mu_hat,
        xlab = "Seconds",
        ylab = "Percent_Outcome",
        level_label = TRUE,
        min.size = 40,
        palette = "Spectral",
        color_level_label = "black")
\end{verbatim}

\begin{figure}[H]
\centering
\includegraphics[width=.8\linewidth]{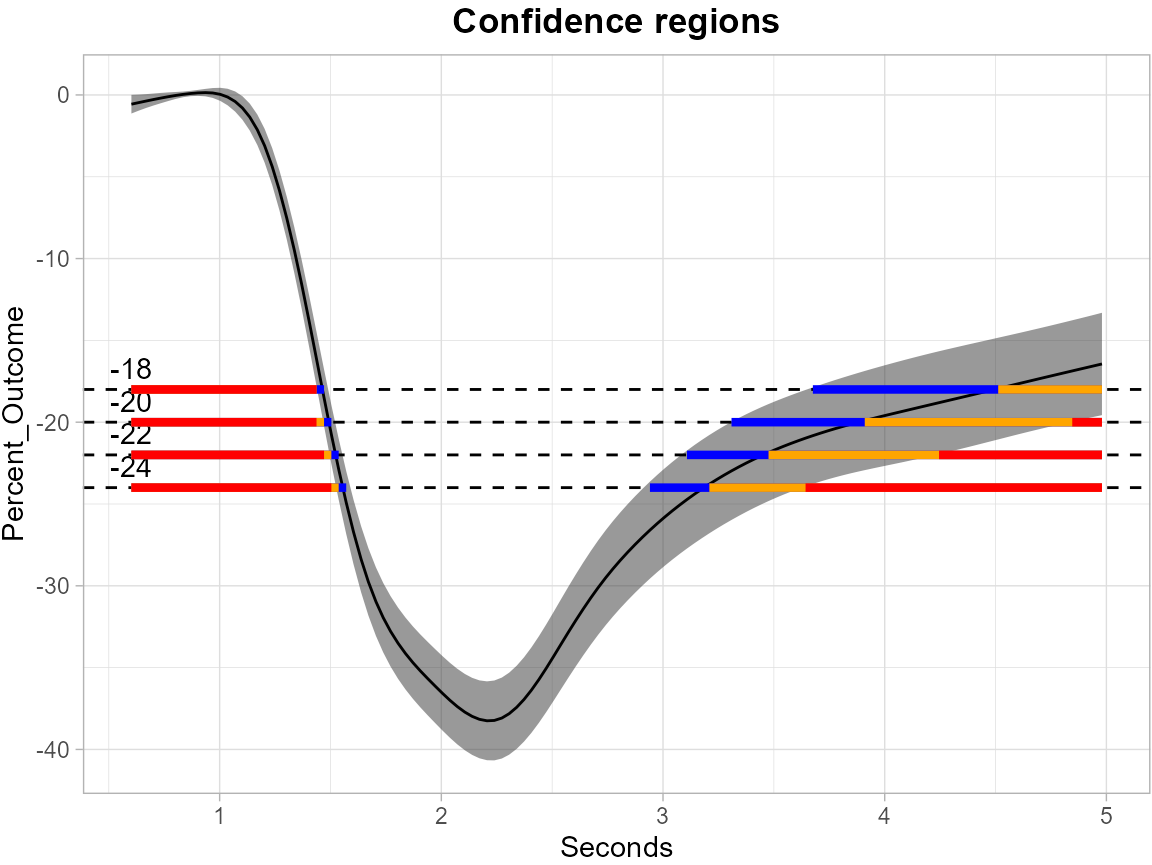}
\end{figure}

  \subsection{SCoRES on Linear/Logistic Regression Model}
 We then look at the application on linear regression models. We will start by using simulated data to construct the
simultaneous outer and inner confidence region from simultaneous
confidence bands (SCB) for the mean outcome using linear regression, with modeling equation as follows:

\[
Y_i = \beta_0 + \beta_1 x_{i1} + \beta_2 x_{i1}^2 + \beta_3 x_{i1}^3
      + \beta_4 x_{i2} + \beta_5 x_{i2}^2 + \beta_6 x_{i2}^3 + \epsilon_i,
\]
where $x_{i,j} \stackrel{\text{i.i.d.}}{\sim} \mathcal{N}(0, 1)$, and $\epsilon_i \sim \mathcal{N}(0, 2)$ for $i=1,\ldots,100$ and $j=1,2$.

We construct simultaneous confidence bands for the expected response surface \(E[\mathbf{Y}|\mathbf{X}_1,\mathbf{X}_2]\), with the predictors \(\mathbf{X}_1\) and \(\mathbf{X}_2\) evaluated on a \(100 \times 100\) grid of equidistant points over the domain \([-1,1]^2\).

\begin{verbatim}
R> set.seed(262)
# generate simulated data
R> x1 <- rnorm(100); x2 <- rnorm(100)
R> epsilon <- rnorm(100,0,sqrt(2))
R> y <- -1 + x1 + 0.5*x1^2 - 1.1*x1^3 - 0.5*x2 + 0.8*x2^2 - 1.1*x2^3 + epsilon
R> df <- data.frame(x1=x1, x2=x2, y=y)
R> grid <- data.frame(x1=seq(-1,1,len=100), x2=seq(-1,1,len=100))
# fit the linear regression model and obtain the SCB for E(y)
R> model <- "y ~ x1 + I(x1^2) + I(x1^3) + x2 + I(x2^2) + I(x2^3)"
R> results <- SCB_linear_outcome(df_fit=df, model=model, grid_df=grid)
\end{verbatim}

The \code{levels = c(-0.3, 0, 0.3)} argument specifies a set of
thresholds, and \code{SCoRES::plot_cs()} function estimates
multiple inverse upper excursion sets corresponding to these thresholds,
and plots the estimated inverse region, the inner confidence region, and
the outer confidence region.

\begin{verbatim}
R> results <- tibble::as_tibble(results)
R> plot_cs(results,
          levels = c(-0.3, 0, 0.3),
          x = seq(-1, 1, length.out = 100),
          mu_hat = results$Mean,
          xlab = "x1", ylab = "y",
          level_label = TRUE, min.size = 40,
          palette = "Spectral", color_level_label = "black")
)
\end{verbatim}

\begin{figure}[H]
\centering
\includegraphics[width=.8\linewidth]{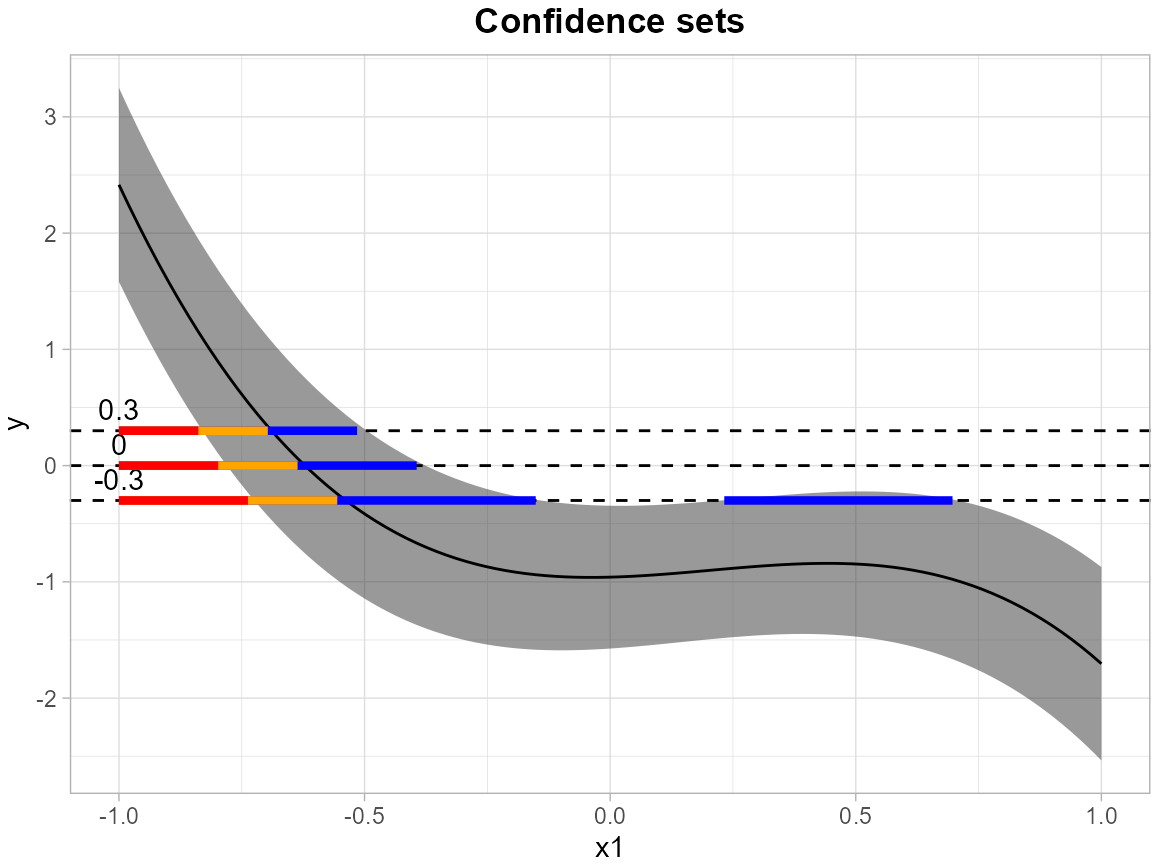}
\end{figure}

In addition to linear regression, \code{SCoRES} also
provides \code{SCoRES::SCB_logistic_outcome()} for estimating the
SCB for outcome of logistic regression.

Here, we use a similar model structure with logistic regression to
estimate the simultaneous confidence band (SCB) for the outcome
probability, with modeling equation as follows:

\[
Y_i \sim Bernoulli(p_i), \qquad
\log\!\left(\frac{p_i}{1-p_i}\right)
 = \beta_0 + \beta_1 x_{i1} + \beta_2 x_{i1}^2 + \beta_3 x_{i1}^3
   + \beta_4 x_{i2} + \beta_5 x_{i2}^2 + \beta_6 x_{i2}^3,
\]
where $x_{i,j} \stackrel{\text{i.i.d.}}{\sim} \mathcal{N}(0, 1)$ for $i=1,\ldots,100$ and $j=1,2$.

We establish simultaneous confidence bands for the expected response
surface \(\text{logit}(E[Y|X_1,X_2])\),
with \(\mathbf{X}_1\) and \(\mathbf{X}_2\) discretized into 100 equidistant points over
the domain \([-1,1]\).

\begin{verbatim}
# generate simulated data
R> x1 <- rnorm(100); x2 <- rnorm(100)
R> mu <- -1 + x1 + 0.5*x1^2 - 1.1*x1^3 - 0.5*x2 + 0.8*x2^2 - 1.1*x2^3
R> p  <- expit(mu)
R> y  <- rbinom(100, size = 1, prob = p)
R> df <- data.frame(x1=x1, x2=x2, y=y)
R> grid <- data.frame(x1=seq(-1,1,len=100), x2=seq(-1,1,len=100))
# fit logistic regression model and obtain SCB
R> model <- "y ~ x1 + I(x1^2) + I(x1^3) + x2 + I(x2^2) + I(x2^3)"
R> results <- SCB_logistic_outcome(df_fit=df, model=model, grid_df=grid)
\end{verbatim}

\begin{verbatim}
R> results <- tibble::as_tibble(results)
R> plot_cs(results,
        levels = c(0.3,0.4,0.5),
        x = seq(-1, 1, length.out = 100),
        mu_hat = results$Mean,
        xlab = "x1", ylab = "y",
        level_label = TRUE, min.size = 40,
        palette = "Spectral", color_level_label = "black")
\end{verbatim}

Likewise, the \code{levels = c(0.3, 0.4, 0.5)} argument
specifies a set of thresholds, and \\
\code{SCoRES::plot_cs()}
function estimates multiple inverse upper excursion sets corresponding
to these thresholds, and plot the estimated inverse region, the inner
confidence region, and the outer confidence region.

\begin{figure}[H]
\centering
\includegraphics[width=.8\linewidth]{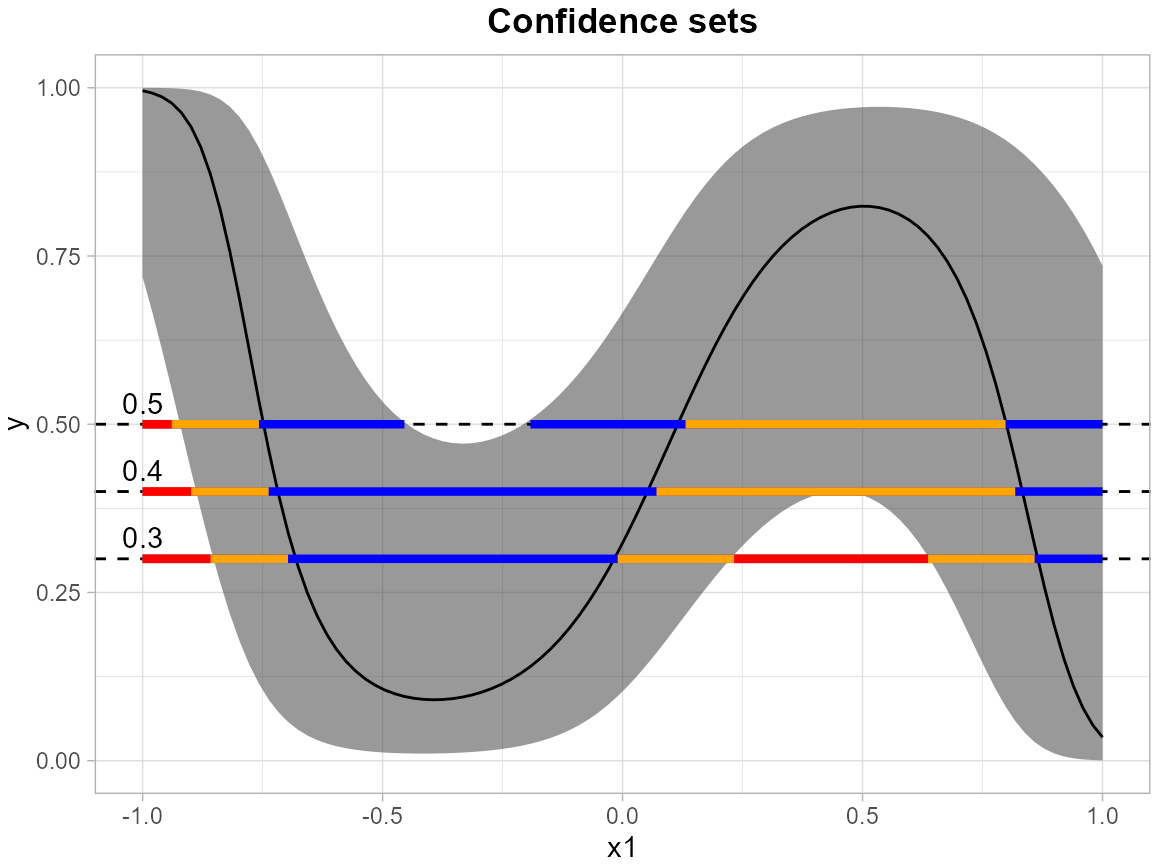}
\end{figure}

Besides, \code{SCoRES::SCB_regression_coef()} can estimate the
SCB for every coefficient in the linear/logistic model. In the following example, we use \code{ipad} dataset from \citep{smith:2023}, which contains tablet-based task performance measures, pupillography features, blood cannabinoid metabolite concentrations, and cardiovascular measures collected 40 minutes after smoking (or after a rest period for controls). 
We establish simultaneous confidence bands for all the coefficients and
the intercept in the model fitted. Here, we set \code{log(t_mmr1)}
as the outcome, and include all variables from
\code{Pupillography}, \code{Tablet (task metrics)} and
\code{Cardiovascular} as predictors.

\begin{verbatim}
R> data(ipad)
R> df <- ipad %>% 
  filter(t_mmr1 > 0) %>% 
  select(p_fpc1, p_fpc2, p_fpc3, p_fpc4, p_fpc5, p_fpc6,
  p_PMC_pctChg, p_auc, t_mmr1,
  i_prop_false_timeout, i_prop_failed1, i_prop_failed2,
  i_judgement_time1, i_judgement_time2, i_time_outside_reticle,
  i_time_on_edge, i_prop_hit, i_correct_reaction2,
  i_reaction_time_max2, i_reaction_time2, i_rep_shapes12,
  i_rep_shapes34, i_memory_time12, i_memory_time34,
  i_composite_score, h_hr, h_dbp, h_sbp, recent_smoke) %>%
  mutate(log_tmmr1 = log(t_mmr1))
df_lin <- df %>% select(-recent_smoke, -t_mmr1)
model_lin <- "log_tmmr1 ~ ."
results <- SCB_regression_coef(df_fit=df_lin, model=model_lin)
\end{verbatim}

The \code{levels = c(0.2, 0.3, 0.4)} argument specifies a set
of thresholds, and \code{SCoRES::plot_cs()} function estimates
multiple inverse upper excursion sets corresponding to these thresholds,
and plot the estimated inverse region, the inner confidence region, and 
the outer confidence region. Here, for illustration, we filter out the
SCB for intercept.

\begin{verbatim}
R> results <- tibble::as_tibble(results[-1,])
R> plot_cs(results,
          levels = c(0.2, 0.3, 0.4),
          x = names(df_lin)[1:(length(names(df_lin))-1)],
          mu_hat = results$Mean,
          xlab = "", ylab = "",
          level_label = TRUE, min.size = 40,
          palette = "Spectral", color_level_label = "black")
)
\end{verbatim}

\begin{figure}[H]
\centering
\includegraphics[width=.8\linewidth]{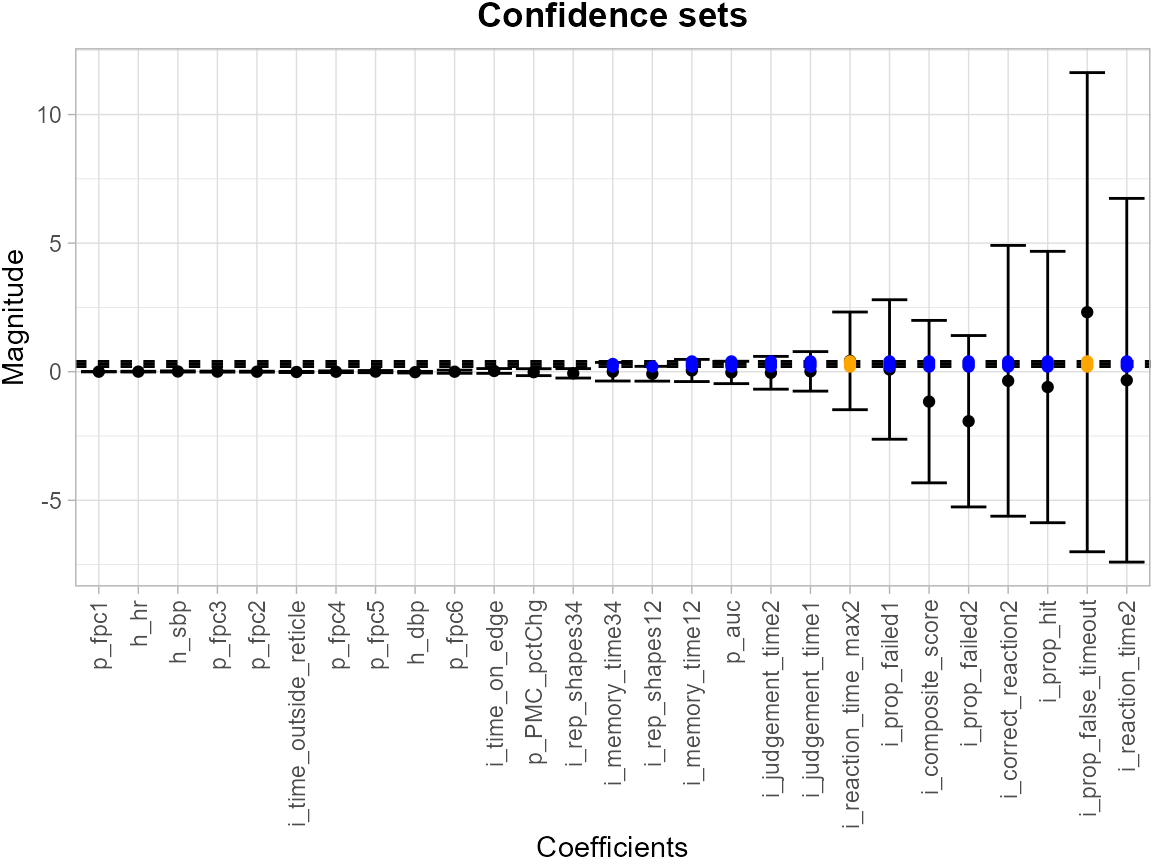}
\end{figure}

We fit a logistic model with \code{recent_smoke} as outcome
with the same predictors in linear regression model, and obtain SCBs for all coefficients and the intercept.

\begin{verbatim}
R> df_log <- df %>% select(-t_mmr1, -log_tmmr1)
R> model_log <- "recent_smoke ~ ."
R> results <- SCB_regression_coef(df_fit=df_log, model=model_log, type="logistic")
\end{verbatim}

The \code{levels = c(2e-07, 3e-07, 4e-07)} argument specifies a
set of thresholds, and \\
\code{SCoRES::plot_cs()} function estimates
multiple inverse upper excursion sets corresponding to these thresholds,
and plot the estimated inverse region, the inner confidence region, and
the outer confidence region. Here, for illustration, we filter out the
SCB for intercept.

\begin{verbatim}
R> results <- tibble::as_tibble(results[-1,])
R> plot_cs(results,
          levels = c(2e-07, 3e-07, 4e-07),
          x = names(df_log)[1:(length(names(df_log))-1)],
          mu_hat = results$Mean,
          xlab = "", ylab = "",
          level_label = TRUE, min.size = 40,
          palette = "Spectral", color_level_label = "black"))
\end{verbatim}

\begin{figure}[H]
\centering
\includegraphics[width=.8\linewidth]{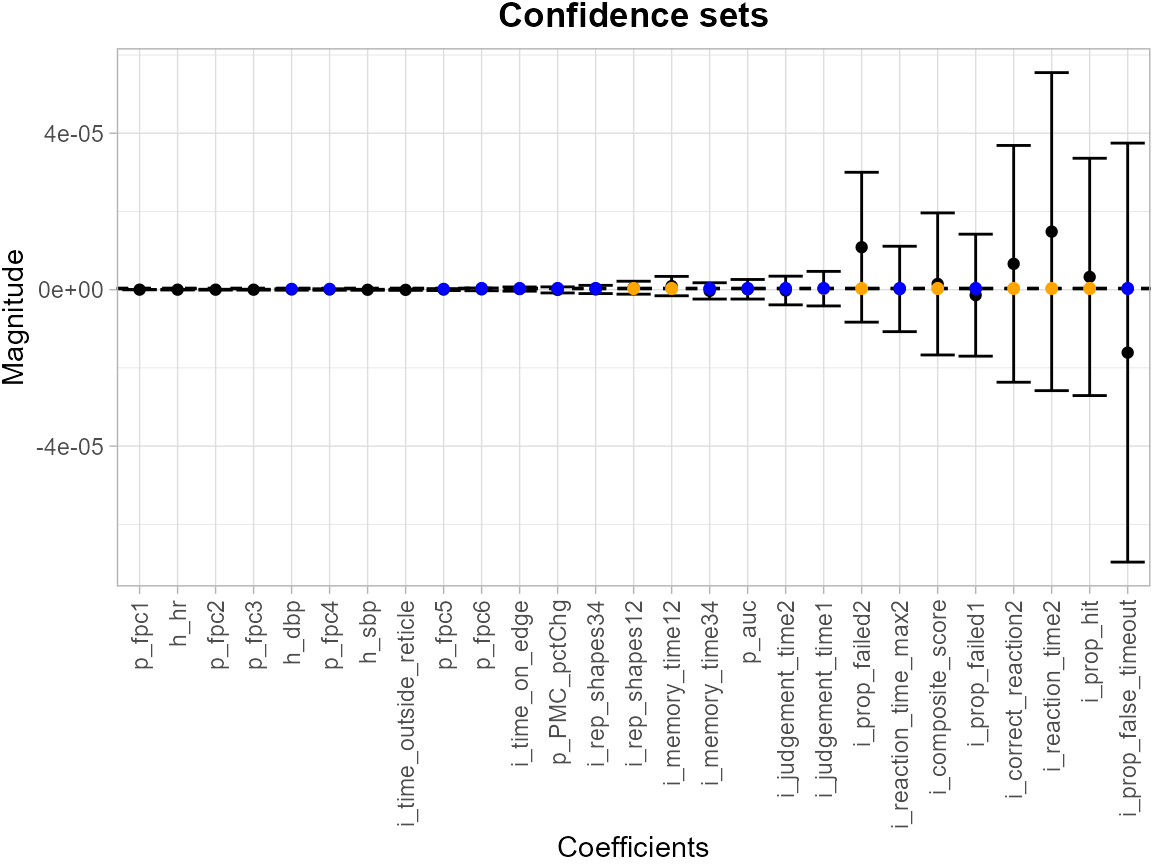}
\end{figure}

  \subsection{SCoRES on Spatial GLS Model}
  Lastly, we illustrate SCoRES application on a spatial GLS model. In the following example, we use climate\_data from SCoRES, which records summer temperature (June–August) in North America
  between the 20th and 21st centuries according to the specific climate model analyzed by \cite{Sommerfeld:2018}. We fit a spatial GLS model for each location with the specification:
\[
\text{Temperature} = \beta_0 + \beta_1 \cdot \text{Group} + \beta_2 \cdot \text{Time}_{\text{current}} + \beta_3 \cdot \text{Time}_{\text{future}} + \epsilon
\]
where:
\begin{itemize}
    \item $\beta_0$: overall intercept
    \item $\beta_1$: coefficient for the group indicator (0 for current years 1971--1999, 1 for future years 2041--2069)
    \item $\beta_2$: coefficient for centered time in current years (set to 0 for future years)
    \item $\beta_3$: coefficient for centered time in future years (set to 0 for current years)
\end{itemize}
More details of the model can be found in the Statistical Methods Section.

We then employ the \code{SCoRES::SCB\_gls\_geospatial()} function to construct SCB for the increase in mean summer temperature across the geographical region. The argument \code{w} specifies the weights for the linear combination of regression coefficients used in SCB construction. Given that the intercept term occupies the second column in \code{data\_fit} and our objective is to estimate the temperature increase, we set \code{w = c(1, 0, 0, 0)}. Additionally, we specify \code{correlation = "corAR1"} to account for temporal autocorrelation in the model residuals.

\begin{verbatim}
# Load data
R> data(climate_data)
# construct confidence region for the increase of the mean temperature 
(June-August) in North America between the 20th and 21st centuries
R> temp = SCB_gls_geospatial(sp_list = climate_data$Z, 
                       level = 2, 
                       data_fit = climate_data$X,
                       w = c(1,0,0,0),
                       correlation = "corAR1",
                       mask = climate_data$mask, 
                       alpha = 0.1)
\end{verbatim}

Heat maps show the estimate of the mean temperature difference. The
first row displays the contours of the outer confidence region,
estimated inverse region, and the inner confidence region, for various
levels. The three plots in the second row display the confidence region
for the inverse sets, where the estimated mean difference is greater or
equal to the individual level 1.5, 2.0, or 2.5 respectively. In the
second row, the blue line is the contour of the outer confidence region,
the green line is the contour of the estimated inverse region and the
red line is the contour of the inner confidence region.

\begin{verbatim}
R> par(mfrow = c(2, 3), mar = c(3, 3, 2, 1)) 
R> p2 = plot_cs(list(scb_up = temp$scb_up, scb_low = temp$scb_low), 
levels = c(1.5, 2,2.5,3), x = temp$x, y = temp$y, mu_hat = 
temp$mu_hat, xlab = "Longitude", ylab = "Latitude", level_label = T, 
min.size = 40, palette = "Spectral", color_level_label = "black")
R> p1 = plot_cs(list(scb_up = temp$scb_up, scb_low = temp$scb_low), 
levels = c(1.5,2,2.5), x = temp$x, y = temp$y, mu_hat = temp$mu_hat, 
xlab = "Longitude", ylab = "Latitude",together = F)
R> p = p2/p1
R> p
\end{verbatim}

\begin{figure}[H]
\centering
\includegraphics[width=\linewidth]{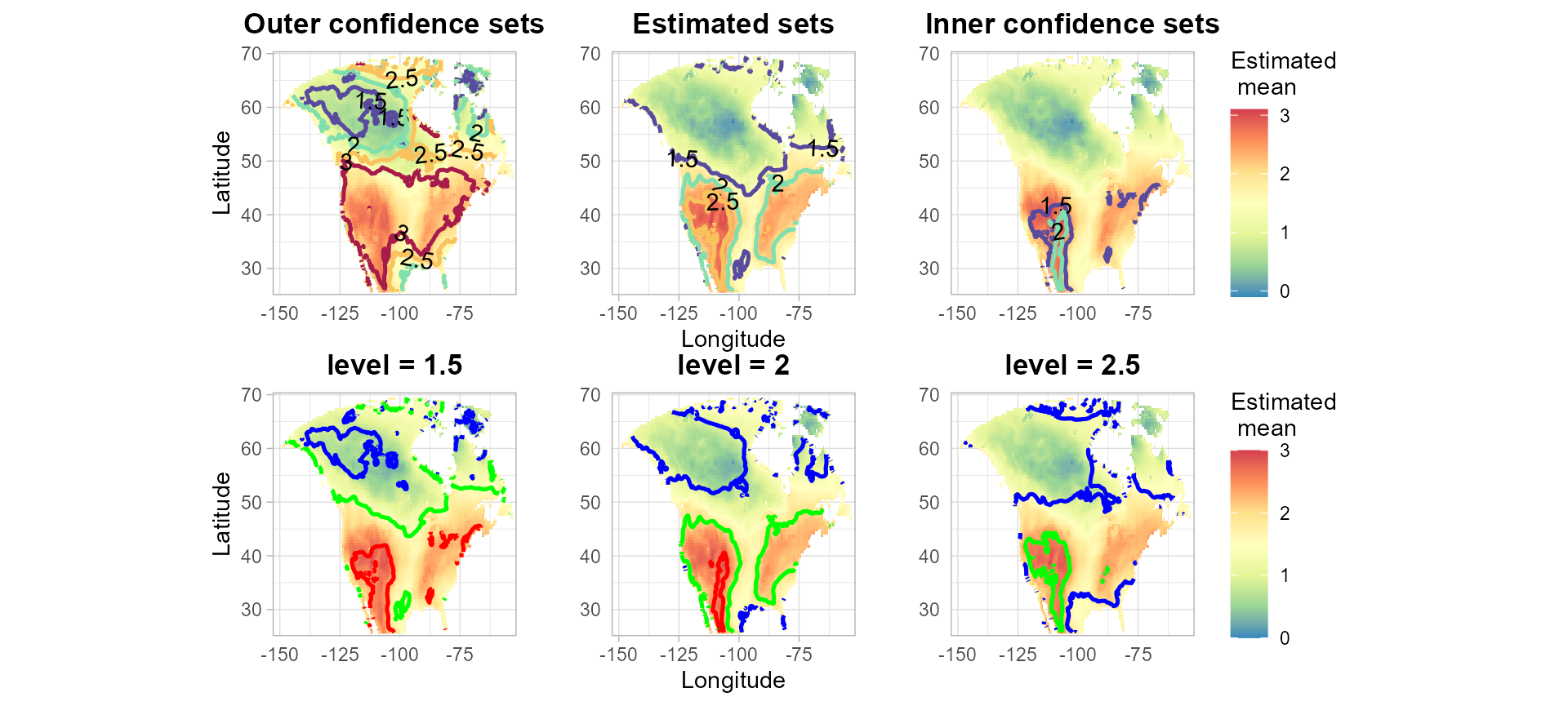}
\end{figure}

  \subsection{SCoRES on Constructing Any Arbitrary SCRs for SCBs}

  Although our package currently supports SCB construction for only the three model types described above, users can also construct SCBs for any arbitrary model on their own. SCoRES provides functionality to estimate the corresponding Simultaneous Confidence Regions (SCRs) based on these user-defined SCBs.

  The following example demonstrates how SCoRES can construct SCRs for the coefficients of a scalar on function (SoFR) model using a pre-constructed SCB.

  We model the binary outcome \textit{use} as a function of the entire pupil-dilation trajectory and several scalar covariates using a functional generalized linear model. Specifically, we fitted a functional logistic regression of the form
\begin{equation}
\label{eq:functional-logistic}
\text{logit}\big[\Pr(\text{use}_i = 1)\big]
= \beta_0
+ \int \beta_1(t) X_i(t)\, dt
+ \gamma_1\,\text{age}_i
+ \gamma_2\,\text{alcohol}_i, \notag
\end{equation}
where $\beta_0$ denotes the intercept term, $X_i(t)$ is the pupil dilation function for subject $i$ measured over time $t$, and $\beta_1(t)$ represents a smooth functional coefficient capturing the time-varying effect of the pupil response on the probability of use. The covariates $\text{age}_i$ and $\text{alcohol}_i$ are included as scalar predictors with corresponding regression coefficients $\gamma_1$ and $\gamma_2$. 

The functional coefficient $\beta_1(t)$ was estimated using cubic regression splines with $k=30$ basis functions, as implemented by the \texttt{lf()} term in the \texttt{refund::pfr()} function. The model was fitted under the binomial family with a logit link, and inference was based on approximate penalized likelihood estimation. The code below illustrates the modeling procedure.

\begin{verbatim}
R> library(tidyverse); library(refund); library(mgcv); library(patchwork)
R> library(SCoRES)
R> data(pupil)

# Prepare data and design matrix for sofr modeling
R> sofr_df = pupil %>%
  select(-percent_change_baseline) %>%
  pivot_wider(names_from = seconds, values_from = percent_change,
              names_prefix = "t_") %>%
  as.data.frame()

R> pupil_mat = sofr_df %>% select(starts_with("t_")) %>% as.matrix()

R> ncols = ncol(pupil_mat)
R> sind = seq(0, 1, len = ncols)
R> smat = matrix(sind, nrow(sofr_df), ncols, byrow = TRUE)

R> sofr_df$smat = I(smat)
R> sofr_df$lmat = I(matrix(1/ncols, nrow(sofr_df), ncols))
R> sofr_df$zlmat = I(sofr_df$lmat * sofr_df$percent_change)
R> gam_use = gam(use ~ s(smat, by=zlmat, bs = "cr", k = 30) + age + alcohol,
              data= sofr_df,
              method = "REML", family = binomial)
\end{verbatim}

Next, we construct SCBs for the coefficient function $\beta_1(t)$ using CMA methods. The code below shows the parametric simulation procedures.

\begin{verbatim}
R> s_pred = seq(0,1, length.out = 100)
R> df_pred = data.frame(smat = s_pred, zlmat = 1, alcohol = 0, age = 0)
# call predict.gam
R> coef_est = predict(gam_use, newdata = df_pred, type = "terms", se.fit = TRUE)

# CMA method for constructing SCBs for beta1(t)
R> lpmat <- predict(gam_use, newdata = df_pred, type = "lpmatrix")
R> inx_beta <- which(grepl("s\\(smat\\):zlmat\\.[0-9]+", dimnames(lpmat)[[2]]))
R> Bmat <- lpmat[,inx_beta]
R> beta_sp <- coef(gam_use)[inx_beta]
R> Vbeta_sp <- vcov(gam_use)[inx_beta,inx_beta]
R> nboot <- 1e4
R> beta_mat_boot <- matrix(NA, nboot, length(s_pred))
R> for(i in 1:nboot){
  beta_sp_i <- MASS::mvrnorm(n = 1, mu = beta_sp, Sigma = Vbeta_sp)
  beta_mat_boot[i,] <- Bmat %*% beta_sp_i
}
R> dvec <- apply(beta_mat_boot, 1, function(x) max(abs(x - beta_hat)/se_beta_hat))
R> Z_global <- quantile(dvec, 0.95)
R> beta_hat_LB_global <- beta_hat - Z_global * se_beta_hat
R> beta_hat_UB_global <- beta_hat + Z_global * se_beta_hat

# Estimate and visualize the corresponding SCR with a threshold of -0.2
R> plot_cs(list(scb_up = beta_hat_UB_global, scb_low = beta_hat_LB_global),
        levels = c(-0.2), x = s_pred,
        mu_hat = beta_hat, xlab = "t_seconds", ylab = "",
        level_label = T, min.size = 40, palette = "Spectral",
        color_level_label = "black")+ylab(expression(beta[1](t)))
\end{verbatim}
The \code{levels = c(-0.2)} argument specifies a single threshold, and \\
\code{SCoRES::plot_cs()} function estimates
the inverse upper excursion set corresponding to the threshold,
and plot the estimated inverse region, the inner confidence region, and
the outer confidence region.

\begin{figure}[H]
\centering
\includegraphics[height=0.3\linewidth, width=0.45\linewidth]{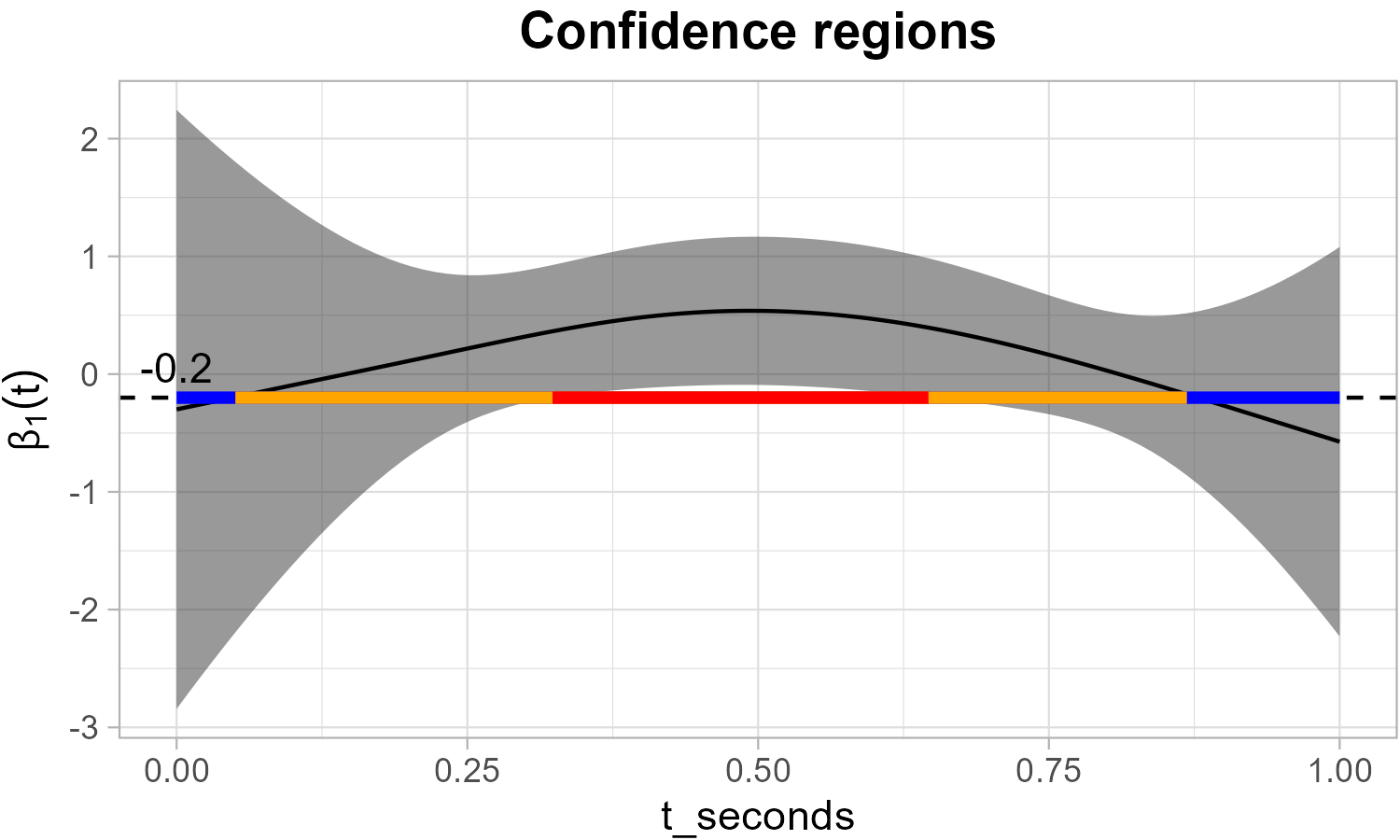}
\end{figure}

\section{Conclusion}
\label{sec:discussion}
The \textbf{SCoRES} R package aims to provide simultaneous confidence bands (SCBs) and inverse confidence regions in a way that intuitively complements common data analysis workflows in R. Similarly to running \texttt{lm()} after fitting a model, a user can simply run functions such as \texttt{SCB\_linear\_outcome()}, \texttt{SCB\_logistic\_outcome()}, or \texttt{SCB\_functional\_outcome()} using their own data to obtain SCBs for target functions or coefficients. The resulting SCBs can then be visualized using \texttt{plot\_cs()}, which also allows the visualization of corresponding simultaneous confidence regions regarding chosen levels. Additionally, the function \texttt{scb\_to\_cs()} can be used to extract estimated regions and their associated simultaneous confidence bounds based on the SCBs and the specified level.

Currently, \textbf{SCoRES} is able to provide SCB estimates for linear, logistic, functional, and spatial regression models. For models that are not currently implemented, users can manually provide their pre-constructed SCBs to \texttt{scb\_to\_cs()} function within \textbf{SCoRES} to convert them into inverse confidence regions and for intuitive visualization.

The \textbf{SCoRES} package implements recent innovations for inverse region estimates with fast construction of confidence regions by simply inverting the SCBs, which does not rely on asymptotic theory or restrictive assumptions. The obtained confidence regions are simultaneous over arbitrary chosen thresholds. With increasing emphasis on interpretable uncertainty quantification in scientific research, the \textbf{SCoRES} package is a user-friendly tool for reporting both function-level inference and inverse region estimates in publications.

\section{Computational Details}
All examples were coded using R version 4.4.2 and \textbf{SCoRES} version 0.1.0. The versions of relevant packages for the examples include: data manipulation with \textbf{dplyr} (1.1.4), \textbf{tidyr} (1.3.1), and \textbf{tibble} (3.2.1); visualization with \textbf{ggplot2} (3.5.1), \textbf{patchwork} (1.3.1), and \textbf{grDevices} (4.4.2); statistical modeling with \textbf{nlme} (3.1-168), \textbf{MASS} (7.3-64), \textbf{refund} (0.1-37), and \textbf{metR} (0.18.1); programming utilities with \textbf{rlang} (1.1.5), \textbf{magrittr} (2.0.3), \textbf{matrixStats} (1.5.0), and \textbf{Matrix} (1.7-3); and base R components including \textbf{stats} (4.4.2), \textbf{utils} (4.4.2), \textbf{forcats} (1.0.0), and \textbf{reshape} (0.8.10).
\section{Acknowledgements}

This research was supported by the National Institute on Drug Abuse (NIDA) of the National Institutes of Health (NIH) under R01DA049800 and P50DA056408, and by the Emory University Rollins School of Public Health Dean's Pilot Innovation Award.


\bibliography{refs}


\end{document}